\documentclass{aa}  

\usepackage{graphicx}
\usepackage{bm}
\usepackage{xcolor}
\usepackage{nccmath}
\usepackage[capitalize]{cleveref}
\crefname{section}{Sect.}{Sects.}
\Crefname{section}{Section}{Sections}

\newcommand{\ev}[1]{\left\langle #1 \right\rangle}

\newcommand{\e}[1]{_{\text{#1}}}

\newcommand{\dd}{\mathrm{d}}

\newcommand{\vect}[1]{\boldsymbol{#1}}

\newcommand{\eps}{\varepsilon}
\usepackage{txfonts}

\newcommand{\codename}{\textsc{Magrathea-Pathfinder}}
\newcommand{\libname}{\textsc{Magrathea}}

\begin{document}

   \title{Magrathea-Pathfinder: A 3D adaptive-mesh code for geodesic ray tracing in $N$-body simulations}
   \titlerunning{\codename{}: an AMR ray-tracing code}

   \author{Michel-Andr\`es Breton\inst{1,2,3} \thanks{\email{breton@ice.csic.es}}
          \and
         Vincent Reverdy\inst{4,5}  \thanks{\email{vince.rev@gmail.com}}
          }

   \institute{Aix Marseille Univ, CNRS, CNES, LAM, Marseille, France
            \and
            Institute of Space Sciences (ICE, CSIC), Campus UAB, Carrer de Can Magrans, s/n, 08193 Barcelona, Spain
            \and
            Institut d’Estudis Espacials de Catalunya (IEEC), Carrer Gran Capit\`a 2-4, 08193 Barcelona, Spain
            \and
            National Center for Supercomputing Applications (NCSA), University of Illinois at Urbana-Champaign, Urbana, Illinois, USA
            \and
            Laboratoire Univers et Théories, Observatoire de Paris, Université PSL, Université de Paris, CNRS, F-92190 Meudon, France
    }
   \date{Received ---; accepted ---}

 
  \abstract{
  We introduce \codename{}, a relativistic ray-tracing framework that can reconstruct the past light cone of observers in cosmological simulations. The code directly computes the 3D trajectory of light rays through the null geodesic equations, with the weak-field limit as its only approximation. This approach offers high levels of versatility while removing the need for many of the standard ray-tracing approximations such as plane-parallel, Born, or multiple-lens. Moreover, the use of adaptive integration steps and interpolation strategies based on adaptive-mesh refinement (AMR) grids allows \codename{} to accurately account for the non-linear regime of structure formation and fully take advantage of the small-scale gravitational clustering. To handle very large $N$-body simulations, the framework has been designed as a high-performance computing post-processing tool relying on a hybrid parallelization that combines MPI tasks with C\texttt{++}11 \texttt{std::thread}s.
  In this paper, we describe how realistic cosmological observables can be computed from numerical simulation using ray-tracing techniques. We discuss in particular the production of simulated catalogues and sky maps that account for all the observational effects considering first-order metric perturbations (such as peculiar velocities, gravitational potential, integrated Sachs-Wolfe, time-delay, and gravitational lensing). We perform convergence tests of our gravitational lensing algorithms and conduct performance benchmarks of the null geodesic integration procedures.
  \codename{} introduces sophisticated ray-tracing tools to make the link between the space of $N$-body simulations and light-cone observables. This should provide new ways of exploring existing cosmological probes and building new ones beyond standard assumptions in order to prepare for the next generation of large-scale structure surveys.
  }


   \maketitle
%

\section{Introduction}
One of the main challenges of modern science is understanding the dark sector of the Universe which dominates its energy content (according to the cosmological concordance model).
Since the discovery of the accelerated expansion of our Universe  \citep{1998Natur.391...51P,1998AJ....116.1009R}, cosmologists have for decades searched for new probes to understand its properties and have successfully confronted the $\Lambda$CDM model with them \citep{planck2018cosmological, Abbott:2018wog, 2018ApJ...859..101S, riess2019large, wong2019holicow, freedman2019CCHP,alam2021completed}. All of these probes share a commonality in that they consist in information coming from photons, using which we can observe the Universe. It is only very recently that we have begun to use other messengers such as gravitational waves to get complementary information \citep{2005ApJ...629...15H,2016JCAP...10..006C}. Nevertheless, photons remain our principal source of information, 
and one can ask how the properties of light modify our perception of the Universe.

The propagation of photons in a non-homogeneous Universe  leads to mainly two effects, which have received a lot of attention over the years and are accounted for to interpret observational data. First, gravitational lensing \citep{schneider1992gravitational,bartelmann2001weak} modifies the apparent position of sources but also alters their observed properties (shape, luminosity) with respect to the case where photons would propagate in a homogeneous Friedmann-Lemaître-Robertson-Walker (FLRW) universe. Second, the position of the emission lines of sources are shifted due to their own motion, which in return leads to an error in estimating distances when assuming a fiducial cosmological model: this is called redshift-space distortions \citep[RSD,][]{kaiser1987clustering,hamilton1992measuring}. The image we have of our Universe is 
therefore altered due to these two phenomena. However, these effects also leave distinct imprints on various cosmological observables, which in turn can help us infer the properties of the Universe. Various current and future missions such as Euclid \citep{laureijs2011euclid} or DESI \citep{desi2016desi} aim at studying these effects through the shape of distant galaxies or their observed spatial distribution.

Due to the increasing quality of data, it is becoming necessary to model the mapping from a statistically homogeneous and isotropic universe to the observed one more accurately. This was done for example regarding the galaxy number counts, accounting for all the effects at first order in metric perturbations within linear theory \citep{yoo2009new,challinor2011linear,bonvin2011what}. However, theoretical prescriptions are limited because they usually cannot properly model the non-linear regime of structure formation. Instead, the use of numerical simulations becomes mandatory to fully understand the clustering of matter, from linear to non-linear scales.

Numerical simulations and, more precisely, dark-matter (DM) $N$-body simulations \citep{hockney1981computer} have been widely used in cosmology to study the large-scale structure of the Universe beyond analytical methods. A lot of work has been done to develop optimised $N$-body codes \citep{kravtsov1997ART,couchman1991mesh, knebe2001MLAPM, teyssier2002cosmological,springel2005cosmological,bryan2014enzo,aubert2015emma,garrison2021abacus}  which allow us to have access to good spatial resolution (small scales) and beat cosmic variance (large scales), while limiting shot noise. Cosmological $N$-body simulations run on supercomputers and use high-performance computing (HPC) techniques, so that today the largest numerical simulations reach trillions of DM particles \citep{potter2017pkdgrav3,heitman2019outerrim,ishiyama2021uchuu}.

$N$-body simulations compute the evolution of the matter density field from initial perturbations at high redshift up to now. However, it is not sufficient to accurately model what we actually observe as one still needs to construct past light cones for some given observers and compute the trajectory of light. This is usually done through the multiple lens formalism \citep{blandford1986fermat}, and ray-tracing post-processing tools have been developed mostly to model the effect of gravitational lensing 
\citep{fluke1999raybundle,jain2000raytracing, fosalba2008onion, hilbert2009raytracing, metcalf2014glamer,giocoli2015disentangling, fabbian2018cmb,gouin2019weak}. However, these methods often compute the lens equation on 2D planes and rely on multiple approximations. Alternatively, several authors have developed ray-tracing algorithms which instead directly compute the geodesic equations in 3D; however the applications are usually still restricted to gravitational lensing \citep{killedar2012gravitational,barreira2016ray} or distance measurements \citep{koksbang2015methods,giblin2016observable}.

Ideally, ray tracing should be general enough to accurately reconstruct the past light cone of an observer and produce a wide range of cosmological observables. This approach has been gaining more and more attention recently \citep{reverdy2014propagation,borzyszkowski2017liger, breton2019imprints,adamek2019bias,lepori2020weak,breton2021theoretical} as it allows for an in-depth study of subtle effects which are currently neglected but could play an important role in future surveys. As evidence of this, \cite{breton2019imprints}  used ray tracing to find the null geodesic connecting an observer to various sources  for the first time in order to estimate the impact of relativistic effects on the clustering of haloes. These authors found that the dipole of the correlation function could be used to probe their gravitational potential in next-generation surveys \citep{saga2022detectability} and therefore test the nature of gravity \citep{bonvin2018testing,saga2021cosmological}.

The goal of this paper is to present the \codename{} framework initially developed in \cite{reverdy2014propagation} and  further developed in several directions since then. It is designed to model the observed Universe as accurately as possible. To this end, we developed ray-tracing techniques that allow us to construct various cosmological observables beyond standard assumptions. In this article we briefly present the basics of \codename{} \citep{reverdy2014propagation} and focus on recent developments and code validation.
In \Cref{sec:methods} we review the main features of our ray-tracing code. In \Cref{sec:tests} we perform convergence tests and conclude in \Cref{sec:conclusion}.

\section{Numerical methods}
\label{sec:methods}
\codename{} is a post-processing numerical framework\footnote{Available at \\ \url{https://github.com/vreverdy/magrathea-pathfinder}} to propagate photons throughout light cones produced by $N$-body astrophysics simulations. The framework is built on top of \libname{} (Multi-processor Adaptive Grid Refinement Analysis for THEoretical Astrophysics), a high-performance library developed to provide highly optimised building blocks for the construction of AMR-based astrophysics applications \citep{reverdy2014propagation}. More specifically, the ray-tracing framework leverages \libname{}'s generic $N$-dimensional hyperoctree abstraction and the associated numerical methods to simplify the handling of the adaptive mesh refinement while ensuring the highest level of performance. Internally, the AMR structure is flattened, each cell being associated with a unique binary index that encodes information about its exact location in the tree. All canonical hyperoctree operations such as finding parents, children, or neighbour cells as well as calculating inter-cell distances are implemented in terms of bit manipulation operations. Each cell is also associated with a data tuple to store the physics quantities attached to a particular location. One of the key features of \libname{} is the heavy use of C\texttt{++} template metaprogramming approaches to guide the compiler through the optimisation process and ensure the highest level of performance regarding the manipulation of indices and physics data \citep{reverdy2015edsl}. In that sense, \libname{} acts as an active library running at compile-time to pre-process \codename{}'s code. The exact implementation of the hyperoctree and low-level numerical algorithms can be found in \cite{reverdy2014propagation} and will be presented in more detail in an upcoming paper focusing on the \libname{} library. For the present paper, we focus on the ray-tracing algorithms in \codename{} made possible by the above-mentioned library.

\subsection{From simulations to \libname{} octree}
\label{subsec:sim_to_magrathea}
As ray-tracing simulations are performed as an independent and subsequent phase of dynamic cosmological simulations, it is possible to leverage the particular geometry of the problem to optimise parallelization schemes and minimise inter-node communications. In practice, the very first step consists in generating a 3D light cone containing all the information relative to the gravitational field and converting it into \libname{}'s 3D-octree data structure. Usually, the light cone is built from an $N$-body simulation. If the simulation uses a particle-mesh (PM) method, then the identification between the cells of the simulations and that of \libname{} can be trivially achieved through the conversion of Cartesian positions into \libname{} indices. The same applies to AMR-based simulations. However, when cosmological simulations rely on another method to compute the gravitational field, one has to first interpolate the gravitational information available at particle locations onto a fixed or adaptive grid that can then be processed by \libname{}'s hyperoctree engine. 

Once the light cone is converted into \libname{}'s octree format, geometrical inconsistencies are checked to ensure that the ray-tracer can work on clean data. One typical problem that may arise during concentric shell extraction happening at the cosmological simulation level is the production of sparse AMR cells at the boundary of two shells. In this case, the geometrical checker adds cells interpolated from a coarser level to build a full tree with either zero or eight children per cell that preserves the original gravitational information. The operation is repeated at every level, from the coarsest to the most refined.

Because the light cone of high-resolution simulations can reach several terabytes of data, it is often not possible to load it on a single node. In this case, a conic domain decomposition of the light cone with overlaps is performed which allows photons to be propagated almost without the need of cross-node communications for the most part of the ray-tracing phase. On top of the conic domain, every computational node gets a copy of a small spherical region centered on the observer to ensure proper independent propagation at very low redshift. The hybrid MPI/C++ thread parallelization approach of \libname{} allows the user to maximise the size of geometrical subdomains and minimise the total size of overlaps while still ensuring a maximal exploitation of computational resources with many threads working at the same time on the same shared memory, each one of them taking care of a particular photon.
Once the octree containing all the relevant gravitational information is built, checked, and distributed across computational nodes, the main ray-tracing phase can begin.

\subsection{Geodesic integration and light propagation}
\label{subsec:geodesic_integration}
\codename{} propagates photons on the null geodesics of the weakly perturbed FLRW metric in Newtonian gauge
\begin{equation}
\dd s^2 = a^2(\eta)\left[-\left(1+2\frac{\Phi}{c^2}\right)c^2\dd\eta^2 + \left(1-2\frac{\Phi}{c^2}\right)\dd\vect{x}^2\right] ,
\label{eq:perturbed_FLRW_metric}
\end{equation}
where $\eta$ and $\bm{x}$ are respectively the conformal time and comoving coordinates, $a(\eta)$ is the scale factor, $\Phi$ the gravitational potential, and $c$ the speed of light. The null geodesic equations are
\begin{eqnarray}
   \label{eq:geodesic_equation1}
              \frac{\textrm{d}^2 \eta}{\textrm{d}\lambda^{2}} &=& -\frac{2a'}{a}\frac{\textrm{d}\eta}{\textrm{d}\lambda}\frac{\textrm{d}\eta}{\textrm{d}\lambda} - \frac{2}{c^2}\frac{\textrm{d}\Phi}{\textrm{d}\lambda}\frac{\textrm{d}\eta}{\textrm{d}\lambda} + 2\frac{\partial\Phi}{\partial\eta}\left(\frac{\textrm{d}\eta}{\textrm{d}\lambda}\right)^2, \\
   \label{eq:geodesic_equation2}
              \frac{\textrm{d}^2 x^i}{\textrm{d}\lambda^2} &=& -\frac{2a'}{a}\frac{\textrm{d}\eta}{\textrm{d}\lambda}\frac{\textrm{d}x^i}{\textrm{d}\lambda} + \frac{2}{c^2}\frac{\textrm{d}\Phi}{\textrm{d}\lambda}\frac{\textrm{d}x^i}{d\lambda} - 2\frac{\partial\Phi}{\partial x^i}\left(\frac{\textrm{d}\eta}{\textrm{d}\lambda}\right)^2,
\end{eqnarray}
where a prime denotes a derivative with respect to conformal time, and $\lambda$ is an affine parameter.
We perform backward ray tracing, meaning that we start the propagation of the photons at the observer today towards the past. The geodesic equations are solved using a fourth-order Runge-Kutta integrator (RK4) with $\mathcal{N}$ steps per AMR cell ($\mathcal{N} = 4$ in common settings), where the photons are initialised using $k^\nu k_\nu = 0$ (with $k^\nu = \dd x^\nu/\dd\lambda$) given the initial direction of the photon $k^i$, and $k^0 = 1$, meaning that at the observer conformal time and affine parameter coincide.

To compute \cref{eq:geodesic_equation1,eq:geodesic_equation2}, we have access to $\eta$, $x^i$, $k^0$ and $k^i$ that are given by the integrator, $a(\eta)$ and $a'(\eta)$ by external pre-computed tables in the fiducial cosmology of the simulation given $\eta$, and $\partial\Phi/\partial x^i$ by the simulation itself. The last subtlety lies in the use of $\dd\Phi/\dd\lambda$ or $\partial\Phi/\partial\eta$, knowing that these terms are simply related by
\begin{equation}
\label{eq:dphidl_1}
\frac{\textrm{d}\Phi}{\textrm{d}\lambda} = \frac{\partial\Phi}{\partial x}\frac{\textrm{d}x}{\textrm{d}\lambda}+\frac{\partial\Phi}{\partial y}\frac{\textrm{d}y}{\textrm{d}\lambda}+\frac{\partial\Phi}{\partial z}\frac{\textrm{d}z}{\textrm{d}\lambda}+\frac{\partial\Phi}{\partial \eta}\frac{\textrm{d}\eta}{\textrm{d}\lambda}.
\end{equation}
Usually, $N$-body simulations do not provide $\partial\Phi/\partial\eta$ and therefore it is easier to rewrite the geodesic equations in terms of $\dd\Phi/\dd\lambda$ only, because it can be estimated by differentiating between two steps of the integration as
\begin{equation}
\label{eq:dphidl_2}
\frac{\textrm{d}\Phi^i}{\textrm{d}\lambda} = \frac{\Phi^{i}-\Phi^{i-1}}{\lambda^{i} - \lambda^{i-1}},
\end{equation}
where the superscripts refer to the integration step. The main drawback of this method is that it strongly depends on the procedure used to build the light cone. For example, if one uses the onion-shell method \citep{fosalba2008onion}, then some artefacts will appear on $\dd\Phi/\dd\lambda$ at the crossing between shells. It is important to note that these subtleties are only important when studying very small effects (such as ISW or time delay).

Ideally, $\partial\Phi/\partial\eta$ should be provided by the simulation. It is the case, for example, with the RayGal simulations \citep{rasera2021raygal} where the authors use a double-layer strategy, meaning that two light cones, slightly shifted in time, are stored. In this case, the time derivative of the potential can be straightforwardly computed for every cell using 
\begin{equation}
\frac{\partial\Phi}{\partial \eta} = \frac{\Phi_2 - \Phi_1}{a_2 - a_1}\frac{\textrm{d}a}{\textrm{d}\eta},
\end{equation}
where the subscripts denote which light cone is used. The main advantage of this method is that it is non-perturbative and therefore accurate even at non-linear scales. Last, we note that in practice $\partial\Phi/\partial\eta$ could also be computed from the density and velocity fields \citep{cai2010fullsky}.

All of the components needed to compute the geodesic equations are available at the location of cells. This means that we need to interpolate them at the photon position at each step (and each sub-step in the RK4 integrator), while accounting for the AMR structure of the light cone. We implemented three different interpolation schemes, namely the nearest-grid point (NGP), cloud-in-cell (CIC), and triangular-shaped cloud (TSC) interpolations, of which the first two were already present in \cite{reverdy2014propagation}. For consistency, the interpolation scheme must be the same as the one used to produce the gravity grid (see \cref{subsec:sim_to_magrathea}). The weighting functions associated with NGP, CIC, and TSC are
\begin{equation}
W\e{NGP}(r_i) = 
\left\lbrace
\begin{array}{lcl}
1  &  \rm{for} & |r_i| < 0.5,  \\
0& \rm{otherwise,} &  
\end{array}\right.
\end{equation}
\begin{equation}
W\e{CIC}(r_i) = 
\left\lbrace
\begin{array}{lcl}
1 - |r_i|  &  \rm{for} & |r_i| < 1.0,  \\
0& \rm{otherwise,} &  
\end{array}\right.
\end{equation}
\begin{equation}
W\e{TSC}(r_i) = 
\left\lbrace
\begin{array}{lcc}
0.75 - r_i^2  &  \rm{for} & |r_i| < 0.5,  \\
\left(1.5-|r_i|\right)^2/2 & \rm{for} & 0.5 < |r_i| < 1.5,  \\
0& \rm{otherwise,} &  
\end{array}\right.
\end{equation}
where $\bm{r}$ is the separation between a cell and the location at which we interpolate, normalised by the cell size. 

\begin{figure}
\includegraphics[width=0.496\columnwidth]{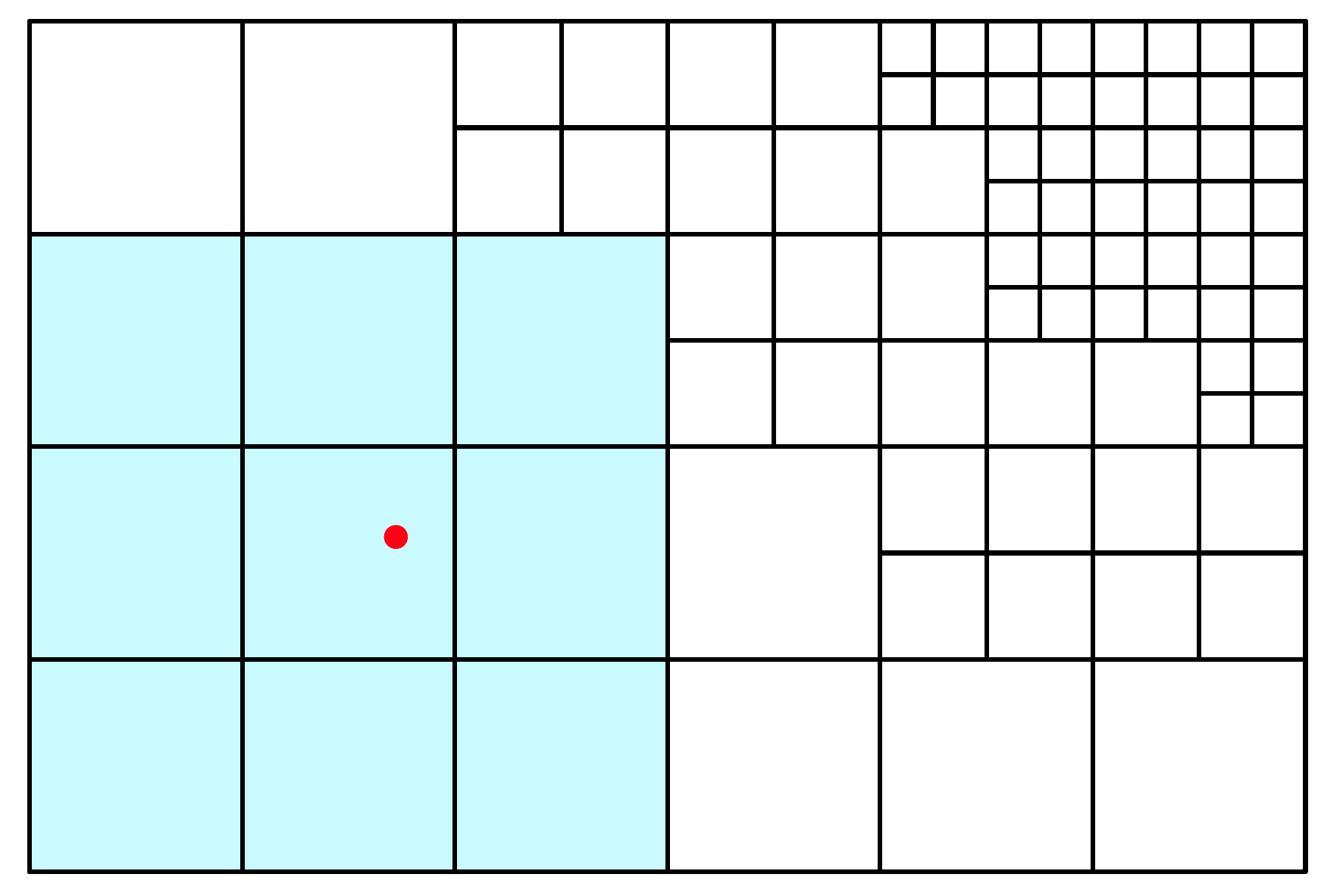}
\includegraphics[width=0.496\columnwidth]{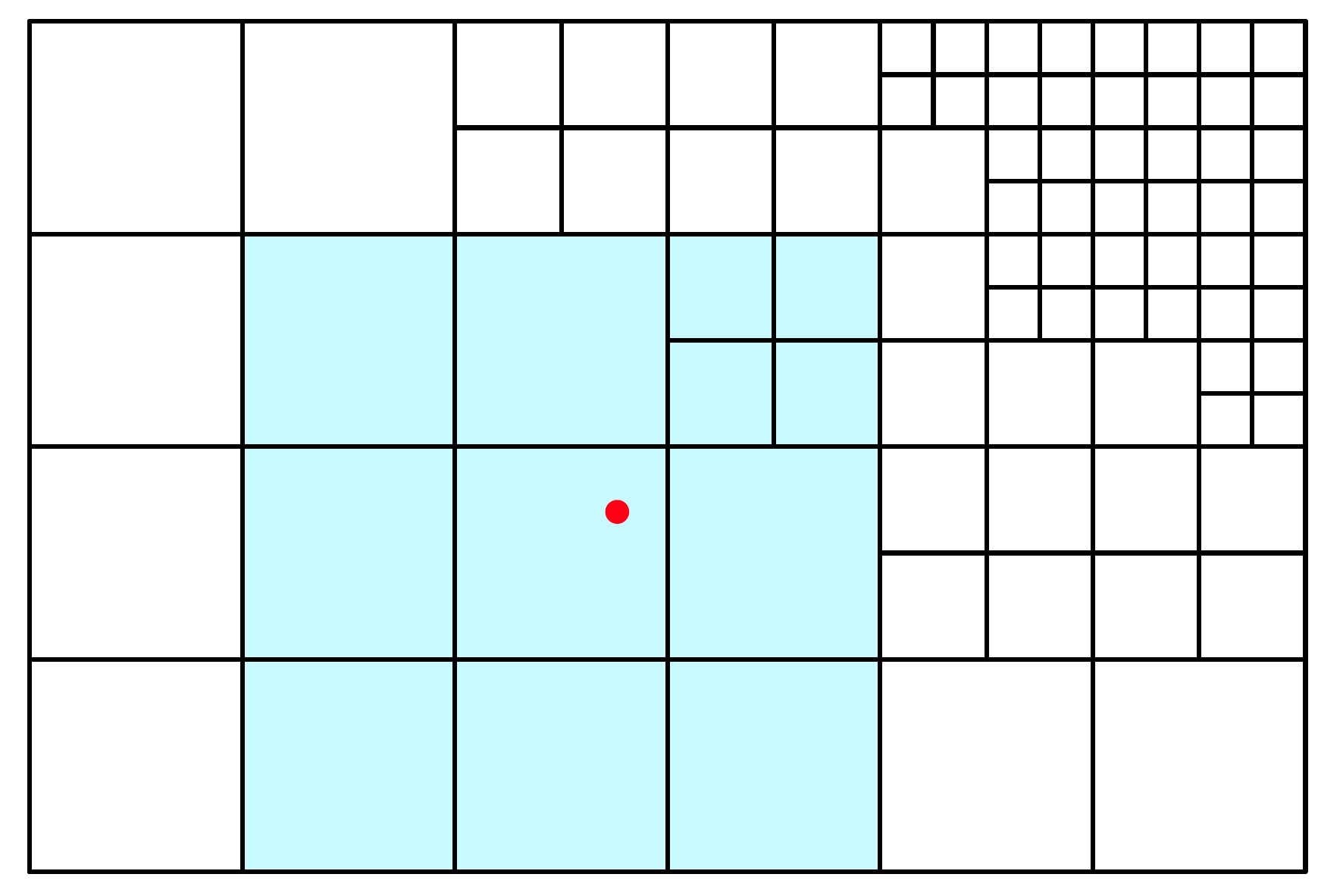} 
\includegraphics[width=0.496\columnwidth]{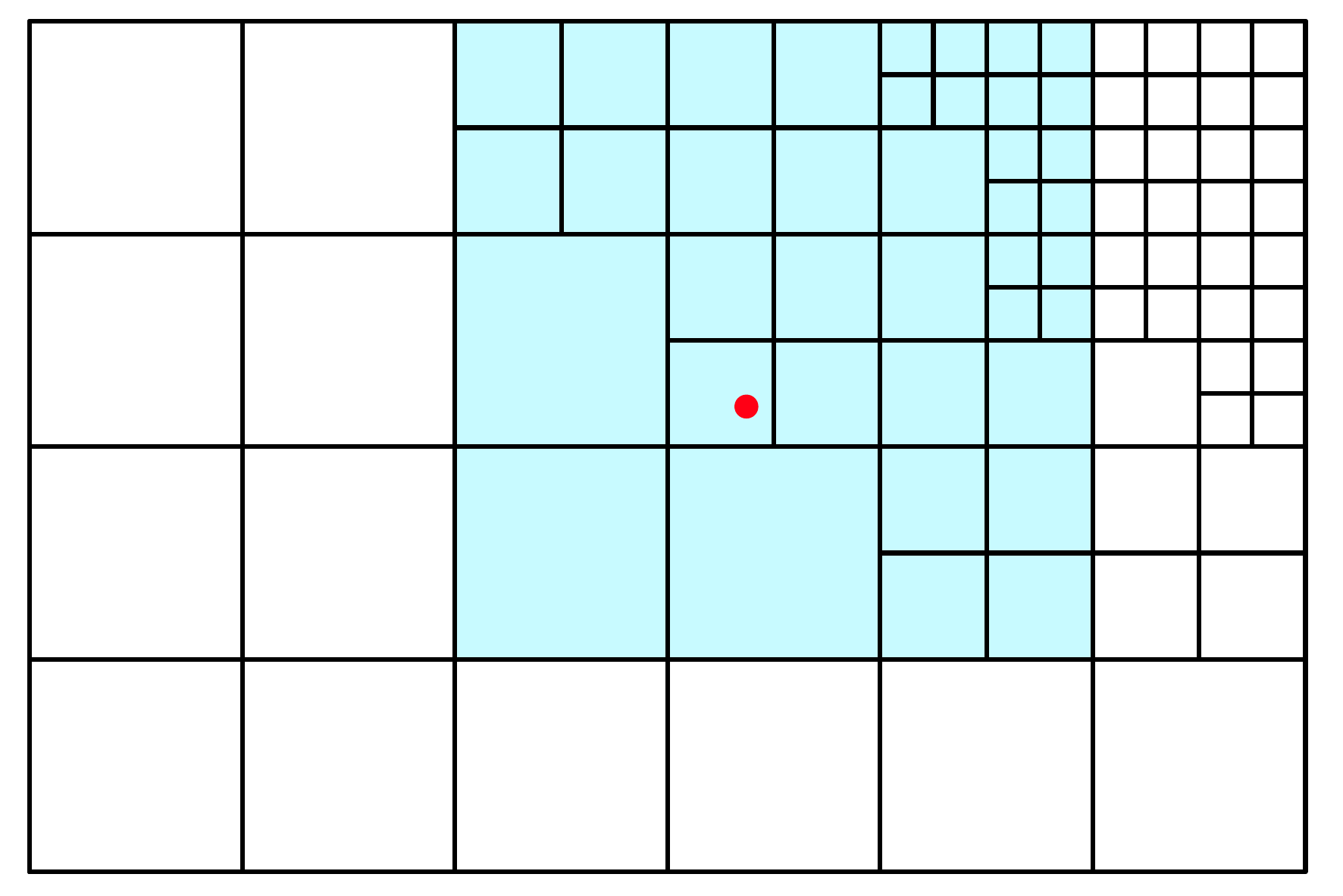} 
\includegraphics[width=0.496\columnwidth]{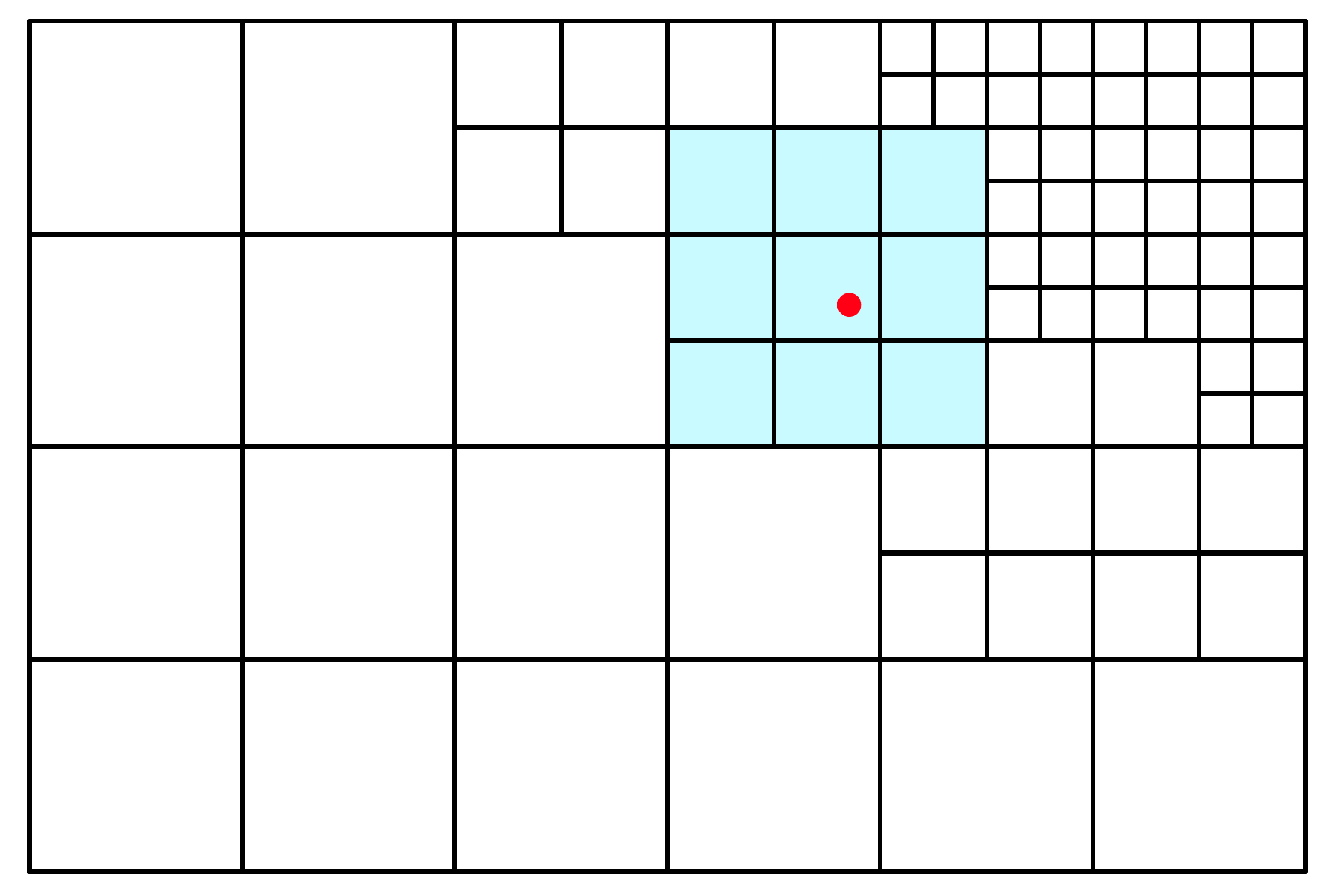} 
\includegraphics[width=0.496\columnwidth]{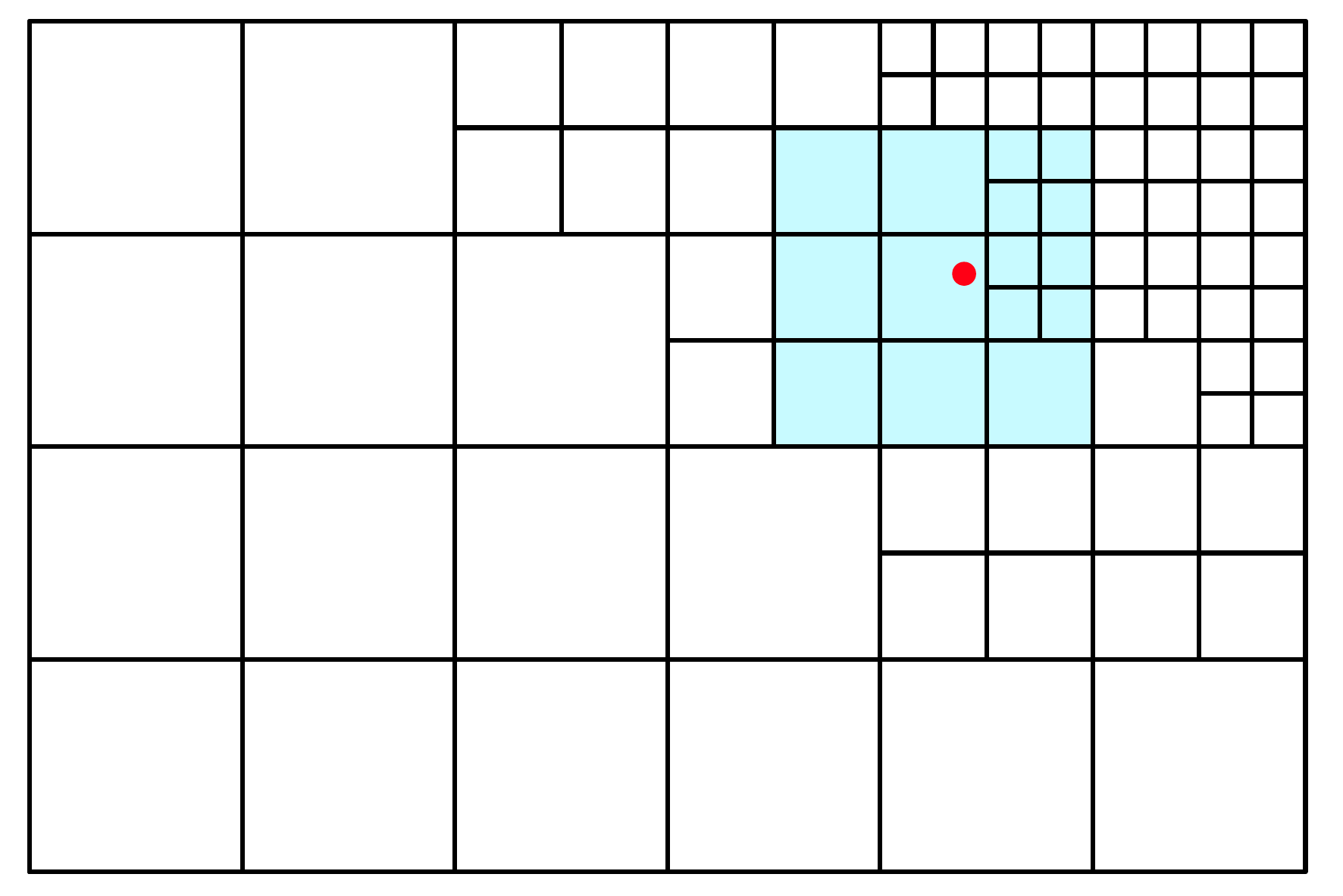} 
\includegraphics[width=0.496\columnwidth]{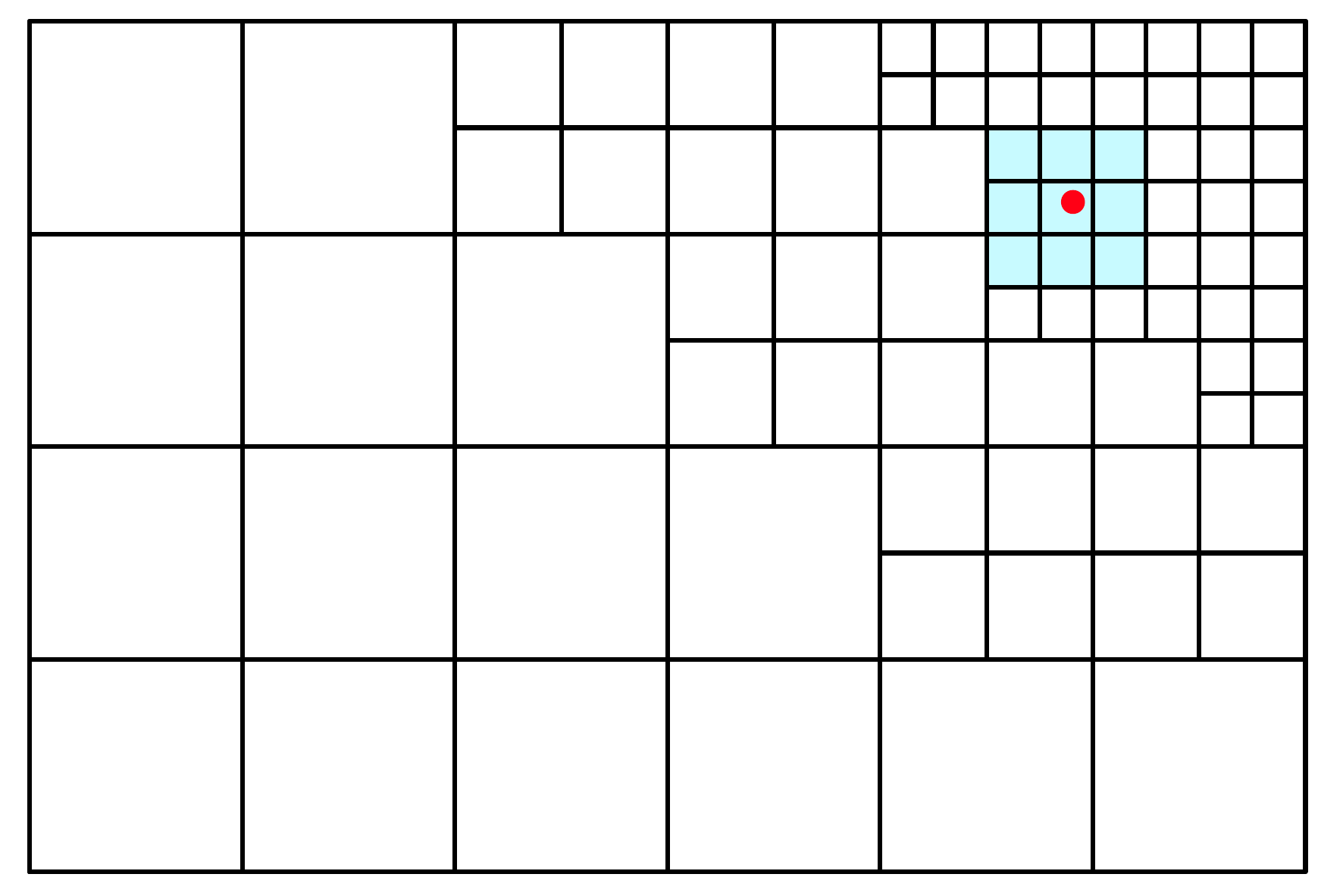} 
    \caption{Illustration of the TSC interpolation scheme at a given location in an AMR grid in 2D. Red points refer to some locations on the grid (which can be for example the position of a photon during its propagation) and the cells used to perform the interpolation are highlighted in blue. The interpolation is done at fixed level, meaning that when the interpolation level is set, we do not use the information from coarser or finer cells.}
    \label{fig:tsc_scheme}
\end{figure}

For NGP, the interpolation is trivial as we take the gravity information from the most refined cell which contains the photon. For CIC and TSC, the interpolation procedure is as follows:
\begin{enumerate}
    \item Estimate the refinement level of the most refined cell containing the photon.
    \item Check if there are 8 (27) neighbouring cells to perform the CIC (TSC) interpolation in three dimensions.
    \item If all the neighbouring cells exist in the octree, compute the interpolation.
    \item If at least one of the neighbouring cells does not exist in the octree, repeat the full procedure at a coarser level. 
    \item If there are not enough neighbours even at coarse level (meaning that we reach the edges of the numerical light cone), the integration is stopped.
\end{enumerate}
We illustrate this procedure in \cref{fig:tsc_scheme}. 
In the top left panel, we see that there are the nine neighbouring cells to perform the 2D TSC interpolation. In this case, the gravity information is easily interpolated at the photon location to compute the geodesic equations. The photon location is updated in the top right panel, where there are also enough neighbours at coarse level. Here, even if the top-right cell also contains more refined cells, they do not contribute to the interpolation and we rather use the coarser cell. In the middle left panel, the photon is in a more refined cell than previously. The code tries to interpolate at this finer level but cannot find the associated neighbouring cells. It therefore performs the interpolation at a coarser level, where all the neighbours exist in the grid. In the middle right panel, the photon is in a new cell at the same level as previously, the only difference being that now there are enough neighbouring cells to perform the interpolation at this finer level. The bottom panels show a similar behaviour as previously but with more refinement. The CIC procedure is similar to that of TSC, except for the fact that in this case we need four neighbouring cells in two dimensions. A nice feature about TSC compared to CIC is that for the latter, the neighbours depend on the position of the photon within a cell, while for the former it is not the case. We take advantage of this by keeping in memory all the neighbours of a given photon at each step, and if the next step of the photon is in the same cell and the TSC interpolation performed at the same level, then we do not have to search again for the neighbouring cells in the octree and therefore gain in performance.
It should be noted that this interpolation procedure does not prevent discontinuities at the crossing between AMR levels; however, we expect this effect to be very faint, especially for the commonly used CIC and TSC interpolation schemes, as we show in \cref{subsec:test_trajectory}.

At each step of the propagation, we keep in memory the step number, scale factor, conformal time, comoving position, affine parameter, redshift, wavevector $k^\nu$, level of the most refined cell, density, gravitational potential and its derivatives with respect to conformal time, affine parameter, and comoving coordinates (i.e. the force). It is also possible to estimate these quantities along the propagation of an unperturbed light ray (this is the so-called `Born approximation') by setting the gravity to zero in the geodesic equations.

Now that we have seen the methodology to propagate photons on null geodesics within the 3D AMR structure of the light cone, we shall turn to the implementation of gravitational lensing and simulated catalogues.

\subsection{Distortion matrix and gravitational lensing}

One direct consequence of light propagation in an inhomogeneous universe is the modification of the apparent position of a source, as well as the modification of its properties (shape and observed luminosity). To derive these properties, we start from the lens equation 
\begin{equation}
    \bm{\beta} = \bm{\theta} - \bm{\alpha},
\end{equation}
where $\bm{\beta}$ and $\bm{\theta}$ are the comoving and observed angular positions on the sky of a source, and $\bm{\alpha}$ the deflection angle. We note that $\bm{\theta}$ is the photon direction at the observer which is directly given by $k^i$ at this location.
The relevant information for weak gravitational lensing is encoded in the Jacobian matrix $\bm{\mathcal{A}}$ which describes the mapping from $\dd^2\bm{\beta}$ to $\dd^2\bm{\theta}$. This gives 
\begin{equation}
  \label{eq:jacobian_matrix}
    \bm{\mathcal{A}} = \frac{\dd^2\bm{\beta}}{\dd^2\bm{\theta}} =
\begin{pmatrix}
     \cos(\omega) & -\sin(\omega) \\
     \sin(\omega)    & \cos(\omega)  
\end{pmatrix} 
\begin{pmatrix}
     1-\kappa-\gamma_1 & -\gamma_2 \\
     -\gamma_2    & 1-\kappa + \gamma_1  
\end{pmatrix},
\end{equation}
where $\kappa$, $\gamma = \gamma_1 + i\gamma_2$ 
, and $\omega$ are respectively the convergence, shear, and rotation of an image. The magnification $\mu$ is the change in observed flux and size of an image (due to the conservation of surface brightness), and is defined as the inverse determinant of $\bm{\mathcal{A}}$.

Usually, ray-tracing codes are designed to compute $\kappa$ and $\gamma$ (and sometimes $\mu$). However, these often make use of several approximations such as Born, plane-parallel, and multiple lens. We now show how to implement the computation of weak-lensing quantities using the 3D propagation on null geodesics described in \cref{subsec:geodesic_integration}. To do so, we use two methods which we refer to as `infinitesimal' and `finite' beams.

\subsubsection{Infinitesimal light beams}
\label{subsubsec:infinitesimal_beams}

First, we consider the usual definition of the distortion matrix, which describes the behaviour of infinitesimal light beams. Formally, \cref{eq:jacobian_matrix} is given by
\begin{equation}
\label{eq:infinitesimal_matrix}
\mathcal{A}_{ab} = \delta_{ab} - \frac{2}{c^2}\int^{\chi\e{s}}_0 \dd \chi \; \frac{(\chi\e{s}-\chi)\chi}{\chi\e{s}}\nabla_a\nabla_b\Phi[\eta(\chi),\vect{x}(\chi)]
\ ,
\end{equation}
where the subscripts refer to the angular coordinates, $\delta_{ab}$ is the Kronecker delta, $\chi$ is the comoving distance of the photon during its propagation, and $\chi\e{s}$ is the distance at the source. We note that here the derivatives are performed along angular spherical coordinates. Because spherical derivatives are not straightforward to compute, we find it easier to first perform derivatives along the 3D Cartesian coordinates, and then rewrite them in terms of spherical ones. At any location on the light cone, we have access to the gravitational field $-\nabla_i\Phi(\bm{x}) \equiv F_i(\bm{x}) = \{F_x(\bm{x}), F_y(\bm{x}), F_z(\bm{x})\}$. It is possible to compute its Cartesian derivatives by differentiating as
\begin{equation}
    \label{eq:laplacian_finite_difference}
    -\nabla_i\nabla_j \Phi(\bm{x}) = \frac{F_j(\bm{x} + h \bm{e}_i)-F_j(\bm{x} - h \bm{e}_i)}{2h},
\end{equation}
where $\bm{e}_i = \{\bm{e}_x, \bm{e}_y, \bm{e}_z\}$ is a unit vector, and $h$ is the derivation step. We found that choosing a derivation step equal to the size of the most refined AMR cell (and therefore highest AMR level) the photon is in, that is $h = 2^{-\rm{level}}$ , is the optimal choice to compute these derivatives (see also \cref{subsec:test_derivation_infinitesimal}). From a numerical perspective, $\nabla_i\nabla_j \neq \nabla_j\nabla_i$ when $i \neq j$. However, the difference is so small that we consider them equal so as to avoid the costly computation of all the possible permutations. The last step is to go from Cartesian to spherical derivatives. Our method is similar to that of \cite{barreira2016ray}, except that we do not resort to the Born approximation and we compute ray-tracing as post-processing. The components of the Laplacian in \cref{eq:infinitesimal_matrix} are given by
\begin{align}
    \nabla_1\nabla_1\Phi &= \sin^2\varphi~ \nabla_x\nabla_x\Phi + \cos^2\varphi~\nabla_y\nabla_y\Phi - \sin 2\varphi ~\nabla_x\nabla_y\Phi, \\
    \nabla_2\nabla_2\Phi &= \cos^2\varphi \cos^2\vartheta~\nabla_x\nabla_x\Phi + \sin^2\varphi\cos^2\vartheta~\nabla_y\nabla_y\Phi 
    \nonumber\\&\quad
    + \sin^2\vartheta~\nabla_z\nabla_z\Phi + \sin 2\varphi\cos^2\vartheta~\nabla_x\nabla_y\Phi
        \nonumber\\&\quad
    -\sin\varphi\sin 2\vartheta~\nabla_y\nabla_z\Phi - \cos\varphi\sin 2\vartheta ~\nabla_x\nabla_z\Phi, \\
    \nabla_1\nabla_2\Phi &= \cos\vartheta\cos\varphi\sin\varphi~(\nabla_y\nabla_y\Phi-\nabla_x\nabla_x\Phi)  
            \nonumber\\&\quad
    + (\cos^2\varphi-\sin^2\varphi)\cos\vartheta~\nabla_x\nabla_y\Phi 
                \nonumber\\&\quad
    + \sin\varphi\sin\vartheta~\nabla_x\nabla_z\Phi-\cos\varphi\sin\vartheta~\nabla_y\nabla_z\Phi,
\end{align}
where $(\varphi, \vartheta)$ is the angular position of the photon in spherical coordinates at each step. Last, we use $\nabla_1\nabla_2\Phi = \nabla_2\nabla_1\Phi$, meaning that we consider that there is no rotation of the image.
We note that the option to use the Born approximation has been implemented by using this method along an FLRW trajectory.

\subsubsection{Finite beams}
\label{subsubsec:finite_beams}
In reality, sources are not infinitesimal but are rather extended. The usual weak-lensing formalism is therefore an approximation of the more accurate finite-beam formalism \citep{fleury2017weak,fleury2019cosmic,fleury2019weak}. In this case we can compute the lensing distortion matrix by launching a beam composed of several close-by light rays which are all integrated on null geodesics independently. The idea to consider several rays to characterise a light beam is similar to the `ray-bundle method' proposed by \cite{fluke1999raybundle} and \cite{fluke2011shape}. First, we launch a reference photon in the direction in which we want to compute the lensing matrix. This photon is used as a reference to know where to stop the beam. For example, we might want to know the lensing quantities at some parameter $p_0$ where $p = \{a, \eta, \chi, z, \lambda\}$ (see \cref{subsec:geodesic_integration}), but we still have the choice to stop the beam light rays at some other parameter $\tilde{p}_0 = \tilde{p}(p_0)$ for the reference photon. We note that all of the parameters are equivalent in an FLRW universe, but this is no longer true when accounting for inhomogeneities.

To compute the distortion matrix, we therefore need to know $p_0$, $\eps$ , which is the beam semi-aperture, and $\bm{\theta} = (\theta_1, \theta_2),  $ which is  the photon direction at the observer. In Cartesian coordinates, the direction of the target is $\hat{\bm{r}} = (\cos\theta_1\sin\theta_2, \sin\theta_1\sin\theta_2, \cos\theta_2)$ where a hat denotes a unit vector. We can define a screen perpendicular to this direction with two orthogonal vectors $\bm{e}_1 = (-\sin\theta_1, \cos\theta_1, 0)$ and $\bm{e}_2 = (-\cos\theta_1\cos\theta_2, -\sin\theta_1\cos\theta_2, \sin\theta_2)$. Now we can launch four rays denoted A, B, C, and D (see also Fig.5 in \citealt{breton2021theoretical}), with initial directions
\begin{eqnarray}
\hat{\bm{r}}\e{A} &=& \hat{\bm{r}} + \tan(\eps)~\bm{e}_1\cdot\bm{u}, \\
\hat{\bm{r}}\e{B} &=& \hat{\bm{r}} - \tan(\eps)~\bm{e}_1\cdot\bm{u}, \\
\hat{\bm{r}}\e{C} &=& \hat{\bm{r}} - \tan(\eps)~\bm{e}_2\cdot\bm{u}, \\
\hat{\bm{r}}\e{D} &=& \hat{\bm{r}} + \tan(\eps)~\bm{e}_2\cdot\bm{u},
\end{eqnarray}
where $\bm{u} = (\bm{e}_x, \bm{e}_y, \bm{e}_z)$. Each ray is propagated on the light cone until $\tilde{p}_0$ so that their final position is given by $\vect{\xi}\e{A}$, $\vect{\xi}\e{B}$, $\vect{\xi}\e{C}$ and $\vect{\xi}\e{D}$. To compute the lensing distortion matrix, we differentiate between the positions of the light rays of the beam. Taking advantage of the fact that the beam is supposed to be small, we can write $\Delta\vect{\beta}=\Delta\vect{\xi}/\chi_0=\hat{\vect{\mathcal{A}}}\,\Delta\vect{\theta}$, where $\chi_0$ is the comoving distance of the reference ray at $p_0$ and $\hat{\vect{\mathcal{A}}}$ the finite-beam distortion matrix. 

A last subtlety is the choice of screen onto which we compute the finite differences. From \cref{eq:jacobian_matrix}, a natural choice for this screen is the one orthogonal to $\bm{\beta} = (\beta_1, \beta_2)$, defined as $\tilde{\bm{e}}_1 = (-\sin\beta_1, \cos\beta_1, 0)$ and $\tilde{\bm{e}}_2 = (-\cos\beta_1\cos\beta_2, -\sin\beta_1\cos\beta_2, \sin\beta_2)$. Alternatively, it is possible to use the more physically motivated Sachs screen, which is orthogonal to the central (reference) photon direction. In this case, the screen is defined with $\tilde{\bm{e}}_1 = (-\sin\zeta_1, \cos\zeta_1, 0)$ and $\tilde{\bm{e}}_2 = (-\cos\zeta_1\cos\zeta_2, -\sin\zeta_1\cos\zeta_2, \sin\zeta_2)$, with $\zeta_1 = \arctan(\hat{k}_y/\hat{k}_x)$ and $\zeta_2 = \arccos(\hat{k}_z)$, where $\hat{\bm{k}} \equiv \hat{\bm{k}}(p_0) = (k_x, k_y, k_z)$ is the direction of the central photon at $p_0$. Finally, the lensing distortion matrix is computed as
\begin{equation}
\hat{\vect{\mathcal{A}}}
\equiv
\frac{1}{2\chi_0\tan(\eps)}
\begin{bmatrix}
(\vect{\xi}\e{A}-\vect{\xi}\e{B})\cdot\tilde{\vect{e}}_1 & (\vect{\xi}\e{C}-\vect{\xi}\e{D})\cdot\tilde{\vect{e}}_1 \\
(\vect{\xi}\e{A}-\vect{\xi}\e{B})\cdot\tilde{\vect{e}}_2 &
(\vect{\xi}\e{C}-\vect{\xi}\e{D})\cdot\tilde{\vect{e}}_2
\end{bmatrix} \ .
\end{equation}
Moreover, instead of stopping the light rays of the beam at $\tilde{p}_0$, we also implemented the possibility to stop them directly on the screen of interest.

Using finite beams to compute the Jacobian matrix allows us to make an accurate treatment of extended sources, which smooths the effect of gravitational lensing on the scale of the beam \citep{fleury2017weak, fleury2019cosmic, fleury2019weak}. This impacts the convergence and shear angular power spectra with respect to the infinitesimal case. A nice agreement between theoretical prediction and numerical estimation was found in \cite{breton2021theoretical} (see also Appendix B for a visualisation of the finite-beam effect on convergence maps).

As this method is purely geometrical and depends on the differential deflection of photons within a beam, we cannot adapt it with the Born approximation. Also, this method does not assume that the off-diagonal terms of the Jacobian matrix are equal, meaning that we have access to the image rotation.
Having implemented the tools to propagate photons and compute the weak gravitational lensing quantities, we now turn to the production of cosmological observables: maps and simulated catalogues.

\subsection{Producing Healpix maps}
\label{subsec:healpix_maps}
First, we consider `direction-averaged' observables \citep{kibble2005average}, which relate to observations in random directions of the sky that are especially relevant for the cosmic microwave background. 
Usually, ray tracing is performed in a pencil beam where all the photons propagate almost in parallel towards the pixels of a plane. In \codename{}, all the photons start from the observer and in this case we find that the most natural frame to homogeneously sample the sky is to use Healpix \citep{gorski2005healpix}. Given a resolution level $N\e{side}$,\footnote{The total number of pixels on the full sky is $N\e{pix} = 12\times N\e{side}^2$} Healpix gives the position of the centre of the pixels. These positions are then used to initialise the photon direction $k^i$ at the observer.

In practice, we first assign the pixels to the different MPI subdomains (see \cref{subsec:sim_to_magrathea}) to ray-trace the different parts of the sky in parallel. Within one MPI task, we use C\texttt{++}11 \texttt{std::thread} multithreading to propagate light rays towards the pixels simultaneously. The user must then specify \texttt{z\_stop\_min} (minimum redshift), \texttt{z\_stop\_max} (maximum redshift), and \texttt{nb\_z\_maps} (number of redshifts at which we compute the maps), which sets (roughly) the redshifts $z_n$ of the output maps. More precisely, the maps are computed at some iso-parameter $p$ surfaces (using the keyword \texttt{stop\_ray}, see also \cref{subsubsec:finite_beams}), and evaluated at $p_n = p(z_n)$, where $p_n$ is the stop criterion computed by launching an FLRW light ray in a very refined homogeneous grid.

Now, we need to specify which quantities we want to estimate. In \texttt{map\_components}, the user can write a list of keywords to output several maps containing the following information:
\begin{itemize}
    \item \texttt{lensing}: The code computes the weak-lensing quantities $\kappa$, $\gamma_1$, $\gamma_2$ and $1/\mu$ using either the infinitesimal method (\texttt{jacobiantype=infinitesimal}, see \cref{subsubsec:infinitesimal_beams}) or the finite-beam method (\texttt{jacobiantype=bundle}, see \cref{subsubsec:finite_beams}). For the latter, an additional map is written which contains the image rotation, and one must specify the stop criterion of bundle (that is, $\tilde{p}$, see \cref{subsubsec:finite_beams}) using \texttt{stop\_bundle}.
    \item \texttt{lensing\_born}: Here we propagate an FLRW light ray in the pixel directions. We then use the infinitesimal method along these trajectories to estimate the weak-lensing quantities.
    \item \texttt{deflection}: The deflection angle, computed as $\bm{\alpha} = \bm{\theta}-\bm{\beta}$, where $\bm{\theta}$ is given by Healpix and $\bm{\beta}$ by the photon position at the map.
    \item \texttt{dens}: The density (computed in the $N$-body solver) interpolated at the iso-$p$ surfaces.
    \item \texttt{dens\_max}: The maximum density probed by the photon during its trajectory until the maps.
    \item \texttt{phi}: The gravitational potential interpolated at the iso-$p$ surfaces.
    \item \texttt{isw}: Integrated Sachs-Wolfe/Rees-Sciama maps.
    \item \texttt{steps}: The number of integration steps for the photon.
    \item Relative differences with respect to their FLRW counterpart for various quantities: $\chi$, $\lambda$, $\eta$, $a$, $z$, with the keywords \texttt{dr}, \texttt{dl}, \texttt{dt}, \texttt{da}, \texttt{dz} respectively.
\end{itemize}

These are the map types currently implemented in \codename{}, and in the future this number will be easily expanded to add more functionalities. The only limitation regarding the number of map components and redshifts at some Healpix resolution is the memory available.

Last, there is one subtlety regarding the computation of the redshift: as we estimate the redshift of the photon at each step of integration, we only perturb the redshift with gravity information (local and integrated terms). Indeed, we do not have access to the velocity which is only available at the position of DM particles (or haloes). However, by adding the compile flag \texttt{-DVELOCITYFIELD}, \codename{} computes the velocity field at the position of the AMR grid by interpolating (using either CIC or TSC) the velocity from all the particles available in the light cone shells around $z_n$ in the subdomain of interest. This means that we need to add data slots on the octree to store $\{v_x, v_y, v_z\}$ at each cell, which produces a heavier octree. This is interesting in particular when we want to compute a map at some constant-redshift surface, where the redshift is notably impacted by the Doppler contributions.

\subsection{Geodesics finder and relativistic catalogues}
\label{subsec:catalogues}
Alternatively, we can produce `source-averaged' observables: these are simulated catalogues which relate to observations at the direction of sources on the sky (such as galaxy or supernovae surveys). As we observe sources thanks to photons that propagated between their emission location and us, we must reproduce the same procedure numerically to construct realistic simulated catalogues. \codename{} already integrates the trajectory of light rays on null geodesic; the only remaining element is to find the appropriate initial condition to link the observer to sources on the light cone.

As described in \cite{breton2019imprints}, we start from the comoving position of a source $\bm{r}$, with comoving angular position $\bm{\beta}$. The goal is to find $\bm{\theta}$ so that the photon angular position at the comoving distance of the source is very close to $\bm{\beta}$. To do so, we iterate over $\bm{\theta}$ and use a Newton-like method to find the null geodesic which connects the observer to the source. This reads
\begin{equation}
\label{eq:newton_method}
    \bm{\theta}_{i+1} =  \bm{\theta}_i - \bm{\mathcal{A}}^{-1} (\bm{\beta}_i - \bm{\beta}),
\end{equation}
where the subscripts refer to the iteration of the root-finding method, and $\bm{\beta}_i$ is the photon angular position at $\chi$ with initial direction at the observer $\bm{\theta}_i$. To avoid any problems due to the system of coordinates, we make all the calculations on a screen orthogonal to $\bm{\theta}_i$. We consider that our method has converged when $|\bm{\beta}_i-\bm{\beta}|< \epsilon$, with $\epsilon$ being the convergence criterion set with the keyword \texttt{cat\_accuracy}. To speed up the calculation, for the first three iterations we impose $\bm{\mathcal{A}} = \mathcal{I}$, with $\mathcal{I}$ being the identity matrix. This should be a good enough approximation when there are no large gravitational fields along the photon trajectory. If the iterations do not converge, we then estimate $\bm{\mathcal{A}}$ with the infinitesimal method (which is faster than the finite-beam one) for two iterations. If convergence is still not achieved, we use a finite-beam method to compute the Jacobian matrix. If after ten iterations convergence is still not achieved (which in our tests represents about one part per million), the sources are saved in some separate files, which we can decide to use if $|\bm{\beta}_i-\bm{\beta}|$ is small enough, or we can run an alternative root finder where we re-run the ray tracing on sources with higher resolution. In this case, we launch photons in the direction of a regular grid centred on $\bm{\theta}_{10}$, and compute $\bm{\theta}_{11}$ from the grid pixel which gives the best agreement. We repeat this process until we achieve the desired accuracy. The size of the grid decreases at every iteration of this procedure. While this last method is slow, it should in the end converge for all the remaining sources. We note that our root-finder algorithm stops whenever one image per source is detected, which corresponds to the weak-lensing regime. For multiply imaged sources, the principal image is likely to be the one to be detected first, and we plan to develop a multiple-image finder in the future to study strong-lensing in more detail.

Finally, for each source we therefore have $\bm{\beta}$, $\bm{\theta}$, $\mathcal{A}$ (computed when $\bm{\theta}$ is known) using either the infinitesimal (with or without the Born approximation) or finite-beam method, as well as various redshifts containing local and integrated terms, depending on the contributions we are interested in. These read
\begin{eqnarray}
\label{eq:firstredshift}
        z_0 &=& \frac{a_0}{a} - 1, \\
        z_1 &=& z_0 + \frac{a_0}{a} \frac{\left[\Phi_o - \Phi_s\right]}{c^2}, \\
        z_2 &=& z_1 + \frac{a_0}{a} \frac{\left[(\bm{v}_s - \bm{v}_o)\cdot\bm{n}\right]}{c}, \\
        z_3 &=& z_2 + \frac{1}{2} \frac{a_0}{a} \frac{\left[|\bm{v_s}|^2-|\bm{v_o}|^2\right]}{c^2}, \\
        z_4 &=& z_3 - \frac{2 a_0}{c^2a} \int^{\eta_o}_{\eta_s} \frac{\partial\Phi}{\partial\eta}\textrm{d}\eta, \\
\label{eq:lastredshift}
    z_5 &=& \frac{(g_{\mu\nu}k^{\mu}u^{\nu})_s}{(g_{\mu\nu}k^{\mu}u^{\nu})_o} - 1, 
\end{eqnarray}
where $z_0$ is the FLRW redshift, and $z_1$ to $z_4$ contain the added contribution of the gravitational potential, Doppler effect, Transverse Doppler effect, and ISW. The scale factor today is given by $a_0$, the subscripts `$o$' and `$s$' refer to evaluations at the observer and at the source respectively, and $g_{\mu\nu}k^{\mu}u^{\nu} = -ak^0 \left[ 1 + \Phi/c^2 + \bm{v}\cdot\bm{n}/c + \frac{1}{2}|\bm{v}|^2/c^2 \right]$. 
In practice, $z_5$ is the true observed redshift (at first order in metric perturbations), while the redshift decomposition from $z_0$ to $z_4$ is particularly interesting to study these effects either in isolation or in combination (see also \citealt{breton2019imprints} for an analysis of these effects on the dipole of the correlation function). For example, by combining $z_0$, $z_1$, and $z_2$ we can infer a measure of redshift that is only perturbed by peculiar velocities, which is the usual framework for RSD studies.
 
 \subsection{Light-ray statistics}
 \label{subsec:light_statsistics}
 Last, we implemented the possibility to propagate light rays and bundles in random directions on the sky, and save several relevant statistics along their trajectories. In each subdomain, the user sets the number of trajectories (i.e. the number of lines of sight). For each trajectory, \codename{} ray traces a light bundle which contains one central ray, and $N$ photons in a circular beam around it. The photons of the beam are evenly spaced on the circle, with semi-aperture set by the user, and each photon propagates on null geodesics independently. From this, \codename{} can either output the full trajectory for all the photons (see \cref{subsec:geodesic_integration} to see which kind of information is saved) or summary statistics for each subdomain or for the full light cone. Saving the full trajectories of all the bundles is interesting for a detailed study of what happens during light propagation at each step of integration, and the bundle method proposed here is more general than that of \cref{subsubsec:finite_beams} which consists in only four surrounding rays. Furthermore, using a spherical bundle with an arbitrary number of photons in principle enables us to study higher order effects of gravitational lensing such as flexion with high accuracy.

\section{Tests and convergence study}
\label{sec:tests}

In this section we describe several tests that we used to check the convergence of the numerical methods described in \cref{sec:methods}. To this end, we used \codename{} on the RayGal simulation\footnote{\url{https://cosmo.obspm.fr/public-datasets/}} \citep{rasera2021raygal} which is based on the PM-AMR RAMSES code \citep{teyssier2002cosmological}. This simulation has evolved $4096^3$ particles (and as many coarse cells) in a (2.625~$h^{-1}$Gpc)$^3$ volume, with the $\Lambda$CDM, WMAP-7 year data best-fit parameters \citep{komatsu2011seven}. The RayGal simulation outputs three light cones using the onion-shell method: a full-sky light cone and two narrow cones with 2500 and 400 deg$^2$ aperture which reach a maximum redshift of $z = 0.5, 2$ and 10 respectively.

\subsection{Performance tests}
\label{subsec:performance_tests}
First, we estimated the run-time performance of \codename{} when propagating photons within the AMR structure of the RayGal light cone. As the MPI and multithreading parallelizations are almost `embarrassingly parallel', that is there is little to no communication between tasks, we expect \codename{} to scale almost linearly with the number of cores. Furthermore, the number of steps per photon depends on the stop criterion, the size of the light cone, and the number of integration steps per AMR cell chosen by the user. Therefore, the relevant quantity to estimate the performances of \codename{} is the time needed to perform a single integration step. This run-time depends on the integrator (we implemented the Euler integration as well as RK4, however only the latter is used) but also on the type of integration. We can identify four types of trajectories, which will give different run times:
\begin{itemize}
    \item FLRW: At any photon location, we assign the gravitational field of an FLRW universe to compute the geodesic equations.
    \item NGP, CIC, TSC: We use the NGP, CIC, and TSC interpolation schemes to estimate the gravitational field from the AMR grid.
\end{itemize}
To perform our tests, we run \codename{} on the French supercomputer Irene, on the Skylake partition composed of Intel Xeon Platinum 8168 processors. We propagate a photon in a given direction of the narrow light-cone of RayGal (with 400 deg$^2$ aperture and maximum redshift $z = 10$) using a single CPU (and single thread). We show the results in \cref{tab:performances}.
\begin{table}
        \centering
        \begin{tabular}{lc} 
                \hline\hline
                Integration & time per step ($\mu$s)\\
                \hline\hline
                FLRW & 0.55\\
                \hline
                NGP & 1.10\\
                \hline
                CIC & 6.0\\
                \hline
                TSC & 14.6\\
                \hline          
        \end{tabular}
        \caption{Ray-tracing run-time for a single integration step, averaged over 100 realisations of the same trajectory and depending on the type of interpolation. We consider a photon trajectory with 4 steps per AMR cell on the narrow light cone of the RayGal simulation until $z = 10$. The total number of integration steps is roughly $5\times 10^4$.}
        \label{tab:performances}
\end{table}
For an FLRW trajectory, \codename{} takes roughly $0.55~\mu $s. As we do not need to estimate the gravitational field, the run-time is mainly that needed to compute the geodesic equations in \cref{eq:geodesic_equation1,eq:geodesic_equation2} with the RK4 integrator. For NGP, we see that it takes 1.10~$\mu$s, which is twice the time of FLRW. The additional time is that needed to obtain the index of the most refined cell the photon is in, find it in the octree, and get the associated data. We note that we search for an index in a sorted vector, meaning that in principle, the larger the octree vector, the longer it takes to find the index (in our case the octree of the subdomain we consider contains roughly $3\times10^8$ elements). When using CIC, \codename{} takes 6~$\mu$s to perform one integration step. It is expected that the CIC interpolation schemes takes more time than NGP, and at first one could have expected $9\times 0.55 = 5~\mu$s to compute the geodesic equations and find the eight neighbouring cells. The slight discrepancy can be explained by the fact that for CIC (and TSC) we need to perform the interpolation at some fixed level, and we loop over the coarser levels if an insufficient number of neighbours are found (see \cref{subsec:geodesic_integration}), which adds some additional run-time. Finally, we see that TSC takes 14.6~$\mu$s, which is less than an optimistic expected run-time of  $28\times 0.55 + 1 = 16.4~\mu$s from the 27 neighbouring cells (instead of 8 for CIC). This comes from the fact that we keep the neighbouring cells in memory, so that we do not need to find them again when the photon is in the same cell and we perform the interpolation at the same level as previously.

Last, we need to estimate how many steps $N\e{tot}$ are needed to reach a given comoving distance $\chi$. Using the $\Lambda$CDM light cones of RayGal, with coarse cell size $x\e{coarse}$ and $n\e{steps}$ the number of steps per AMR cell (set by the user), we find that the total number of integration steps as a function of the distance is well fitted by
\begin{equation}
    N\e{tot}(\chi) \approx \left[1.33 -2.37\times 10^{-5} ~\raisebox{0.4ex}{$\chi$}\right] \frac{n\e{steps}}{x\e{coarse}}\chi,
    \label{eq:ntot_steps_distance}
\end{equation}
between $z = 0.1$ and $z = 10$, where $\chi$ and $x\e{coarse}$ are in units of $h^{-1}$Mpc. If there was no AMR, we would expect $N\e{tot}(\chi) = n\e{steps}~\raisebox{0.3ex}{$\chi$}/x\e{coarse}$, which is different from \cref{eq:ntot_steps_distance}. This difference comes from grid refinement (AMR), especially at low redshift where the late-time small-scale clustering is more important.

\subsection{Accuracy of the interpolation schemes}
\label{subsec:test_trajectory}

One of the main features of AMR is the fact that we have access to very non-linear scales of structure formation in high-density regions. As our ray-tracing procedure adapts to the AMR level of the simulation light cone, we might wonder how well we recover the gravitational potential in these regions, and how it depends on the interpolation schemes described in \cref{subsec:geodesic_integration}. We use the methods in \cref{subsec:light_statsistics} to save the full trajectory of a single light ray for a given line of sight, with the three interpolation schemes previously described by setting the compile option, \texttt{-DORDER =} 0, 1, and 2 for NGP, CIC, and TSC, respectively. We note that RayGal uses a TSC version of \textsc{Ramses} to compute the potential, which means that for consistency we should use the same interpolation scheme. However, it can be interesting to visualise the differences between these different types of interpolation and how it behaves with AMR.

\begin{figure*}
\includegraphics[width=\columnwidth]{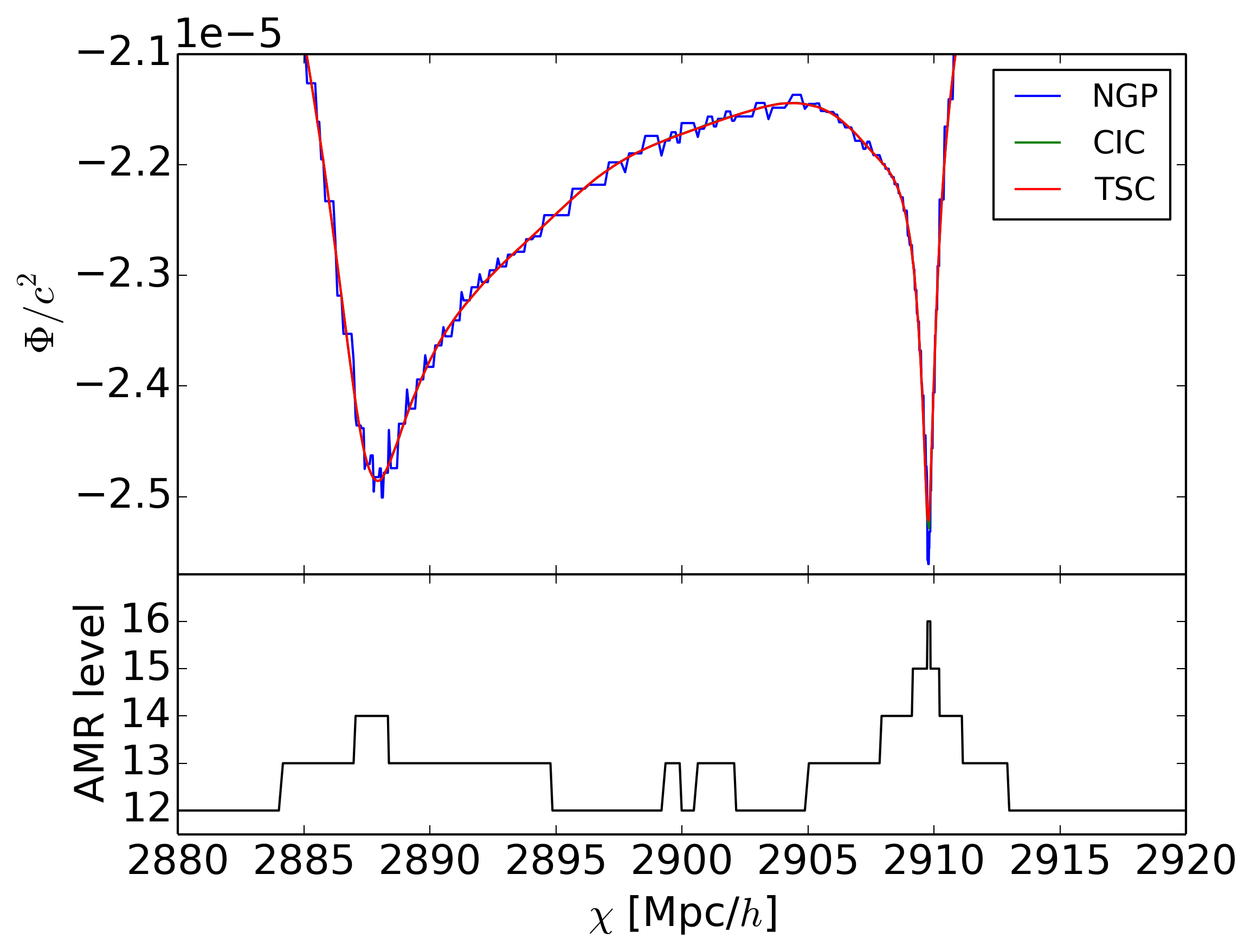}
\includegraphics[width=\columnwidth]{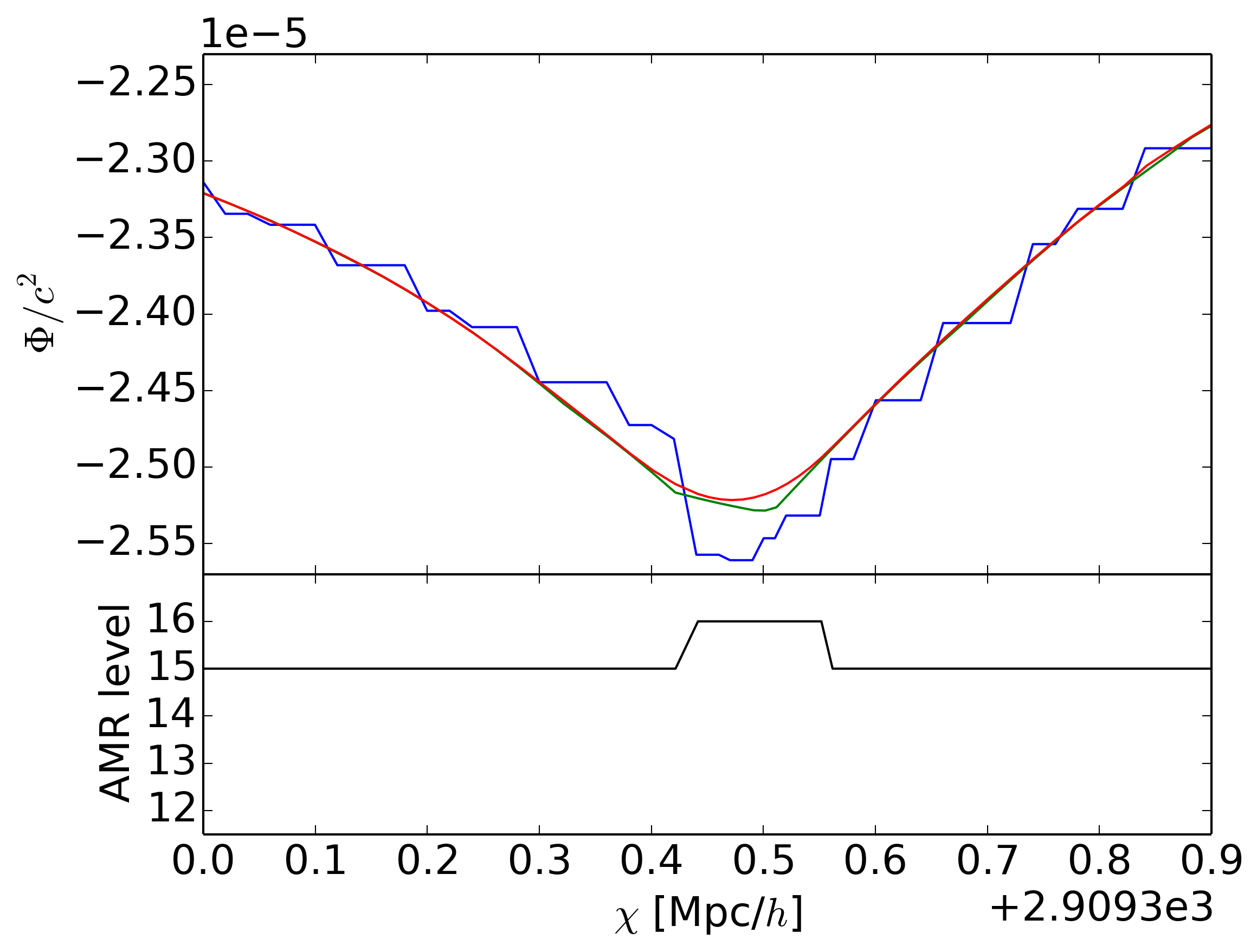} 
    \caption{Gravitational potential and AMR level along a photon trajectory, as a function of the comoving distance for different interpolation schemes. The right panel is a zoom onto the second potential well seen in the left panel around 2910$~h^{-1}$Mpc. The potential minima coincide with a higher clustering of matter, which refines the AMR grid (as seen in the subplots).}
    \label{fig:potential_amr}
\end{figure*}

In \cref{fig:potential_amr}, we show the gravitational potential at the photon position for each integration step as a function of the comoving distance to the observer. 
Additionally, we also show the AMR level of the most refined cell at the photon location. In the left panel, we see two potential wells around 2887 and 2910 $h^{-1}$Mpc, which is roughly $z\approx 1.3$ in the fiducial cosmology of the simulation. These potential wells reach roughly $-2.5\times 10^{-5}$, which is slightly less than 10~km/s (the order of the gravitational redshift for galaxies and cluster). We can clearly see the difference between NGP and the other two interpolation schemes, as the former shows some sharp `jumps' when the photon goes in a different cell while the others seem to exhibit a much smoother behaviour. At the same time, we see that the AMR level follows a similar pattern as the gravitational potential, starting from level 12 (which is the coarse level), and increases when the photon enters the potential wells. This is logical, because these potential wells come from small-scale clustering of matter, which also causes the $N$-body solver to refine the grid in these high-density regions. In particular, we see that for the second, deeper, and sharper potential well, the AMR level goes to 16 (while for the first well it only reaches level 14). The right panel shows a zoom of the second potential well in order to better visualise the differences between the interpolation schemes in these high-density regions. We indeed see the `stair' behaviour of the NGP interpolation, while the CIC interpolation is very smooth. However, we can see that CIC seems to give a potential which linearly evolves by parts along the trajectory. This is expected, because CIC is a tri-linear interpolation and we use four steps per AMR cell. This means that the value of a field estimated by the same eight neighbours of the CIC scheme necessarily evolves linearly along any direction. Last, we see that the TSC scheme agrees very well with CIC far from the trough, while at the minimum of the potential it seems much smoother and would give better results when estimating its derivatives (which is important because weak lensing is sensitive to the second derivative of the gravitational potential) compared to CIC. This is also expected because TSC is higher order.

As a final note, we can see that although our method does not necessarily prevent discontinuities at the crossing between AMR levels, we do not see sharp cuts in the gravitational potential when using either CIC or TSC. This suggests that our methodology should give good results even for very small scales.

\subsection{The importance of AMR}
\begin{figure}
\includegraphics[width=\columnwidth]{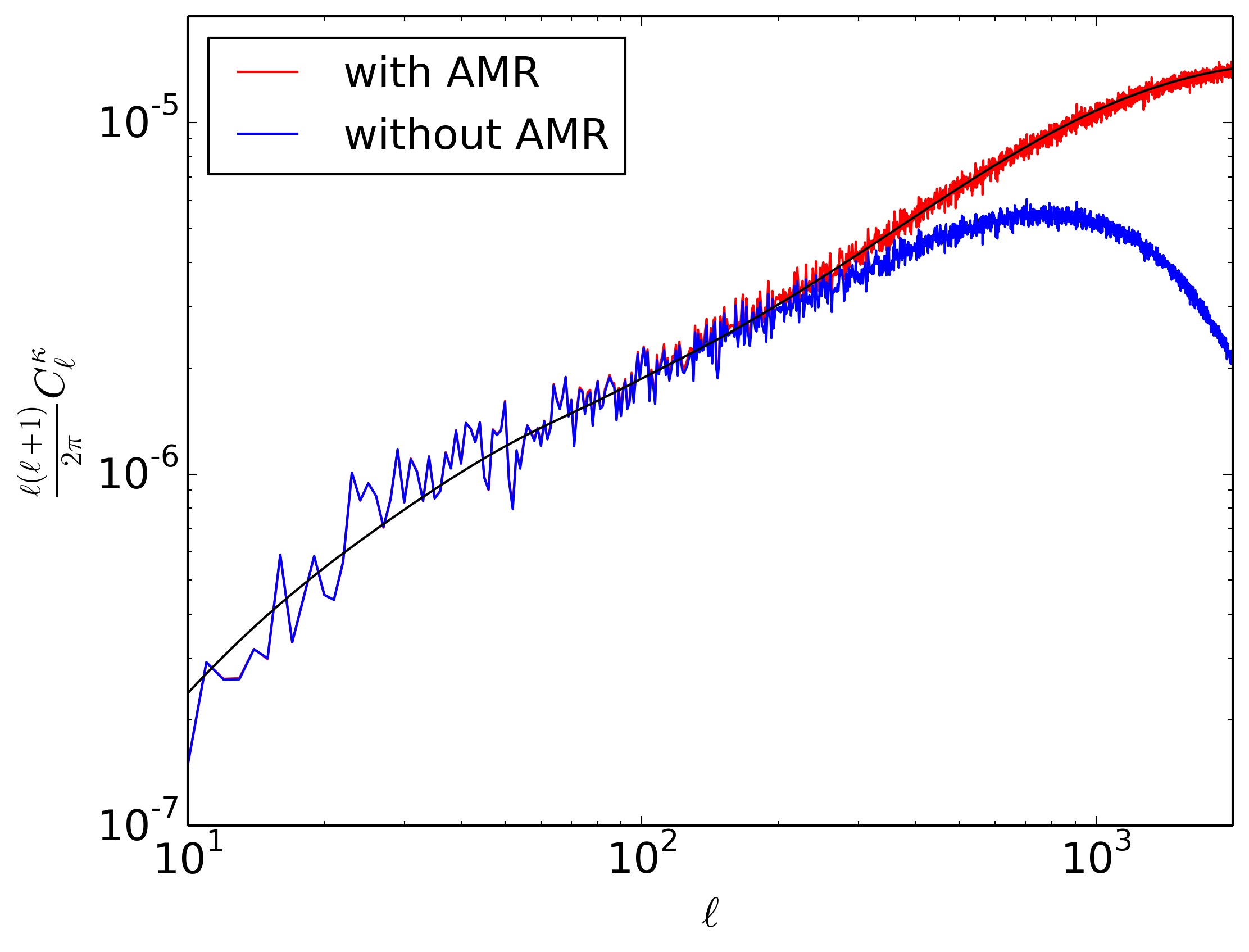} 
    \caption{Convergence power spectra with AMR (red) and without (blue) at $z = 0.45$, using Healpix with $N\e{side} = 2048$. The black line is the theoretical prediction computed using \textsc{Nicaea} \citep{kilbinger2017precision} with \textsc{Halofit} \citep{smith2003stable} parameters fitted on our simulation. When AMR is accounted for, we have a nice agreement between the numerical results and theoretical prediction. When it is not, the power spectrum experiences a damping at small angular scales (high $\ell$), where the discrepancy becomes large already at $\ell \approx 300$.}
    \label{fig:convergencePS_AMR}
\end{figure}

Because most of the gravitational lensing power lies in small scales, we already expect AMR to be an important aspect of the ray-tracing procedure. In \cref{fig:convergencePS_AMR}, we quantify its impact through the estimation of the angular power spectrum of the convergence computed with \textsc{anafast} on the full-sky light cone at $z = 0.45$, between $\ell\e{min} = 10$ and $\ell\e{max} = 2000$ (because in that case $\ell\e{max} \sim N\e{side}$, which should give sufficient accuracy). 
First, we note that the angular power spectrum computed with AMR is in excellent agreement with the theoretical prediction. This validates our methodology (in the present case, that of the finite-beam method) to compute the lensing distortion matrix. Second, we see that the angular power spectrum without AMR (i.e. we only consider the coarse level for the ray tracing) departs early (around $\ell \sim 200-300$) from the AMR case and exhibits a strong damping at small scales. A similar trend was noted by \cite{lepori2020weak}. This shows that AMR is extremely important for  recovering the correct statistical properties of gravitational lensing when modelling light propagation in a PM $N$-body simulation.

\subsection{Propagation and number of steps per AMR cell}
\label{subsec:nsteps_per_amr_cell}
We now verify the appropriateness of our choice of making four integration steps per AMR cell. It was first shown in \cite{reverdy2014propagation} that using four steps was ideal for correctly recovering the redshift up to double precision with respect to an analytical calculation when propagating an FLRW light ray with an RK4 integrator until $z = 25$ in the DEUS-Full Universe Run simulations \citep{alimi2012deus, rasera2014cosmic,bouillot2015probing}. However, it is unclear whether or not this choice is still relevant when accounting for an inhomogeneous universe.

\begin{figure}
\includegraphics[width=\columnwidth]{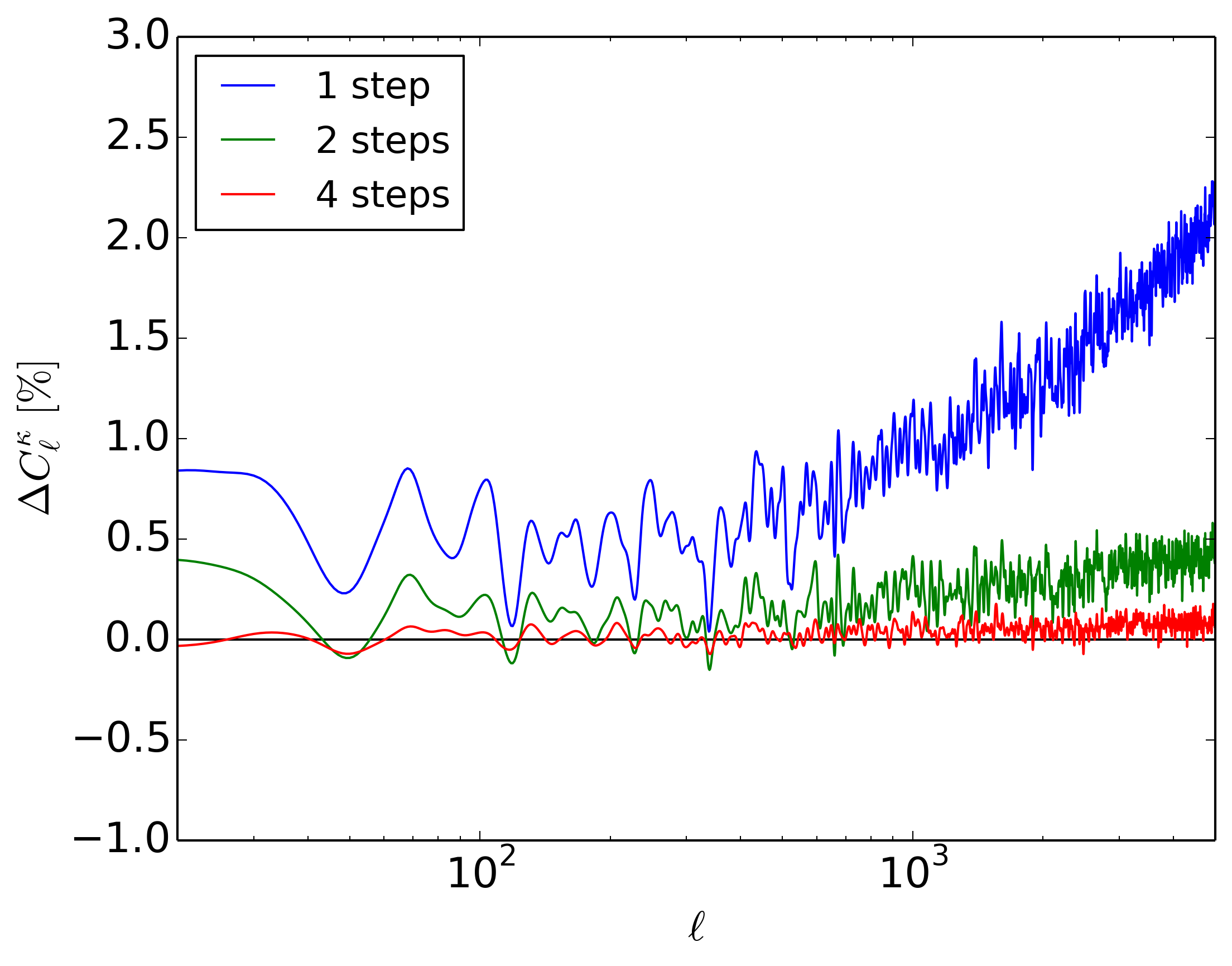} 
    \caption{Relative difference in the angular power spectrum of the convergence as a function of the number of steps per AMR cell during the propagation. The reference for the relative difference is taken as the angular power spectrum with eight steps per AMR cell. The convergence is computed using an infinitesimal method on a Healpix map with $N\e{side} = 4096$ at $z = 1.5$. We see that we achieve numerical convergence when using four steps per AMR cell.}
    \label{fig:reldiff_steps_per_cell}
\end{figure}

In \cref{fig:reldiff_steps_per_cell}, we show the impact of taking one, two, or four steps on the convergence angular power spectrum with respect to the very conservative case where we use eight steps per AMR cell.
To do so, we used a narrow light cone with 2500 deg$^2$ aperture, and produced Healpix maps (see \cref{subsec:healpix_maps}) of the convergence at $z = 1.5$ with different values of \texttt{nsteps}. To correctly estimate the power spectrum with a mask we use PolSpice \citep{szapudi2001fast,chon2004fast}.
We see that the effect here is very small (at most a few percents). When taking one step per AMR cell, we see a roughly $0.5\%$ bias at large angular scales, which increases up to $2\%$ at $\ell = 5000$. For two steps, the angular power spectrum departs from the reference one around $\ell \approx 1000$ to reach $0.5\%$ at most. Finally, we find that the convergence angular power spectrum with four steps is indistinguishable from its eight-steps counterparts. This is further evidence that there is no need to use more than four steps per AMR cell to propagate light rays.

We note that if subpercent precision is not needed, then two steps per AMR cell is sufficient. This should decrease the run-time by a factor of two and hence be very interesting for HPC. In any case, the user can decide to use an arbitrary number of steps by using the keyword \texttt{nsteps}.

\subsection{Infinitesimal case, choice of the derivation step}
\label{subsec:test_derivation_infinitesimal}

Now we turn to the estimation of the lensing distortion matrix with the infinitesimal method as described in \cref{subsubsec:infinitesimal_beams}. The reason we only study the convergence of the infinitesimal method and not that of the finite-beam one is that the former depends on an arbitrarily chosen derivation step, while the latter does not depend on any arbitrary choice (except for the beam aperture at the observer, which is a parameter set by the user and is physically motivated, with its impact clearly understood from a theoretical perspective; see \citet{fleury2019cosmic} and \citet{breton2021theoretical}. 

\begin{figure}
\includegraphics[width=\columnwidth]{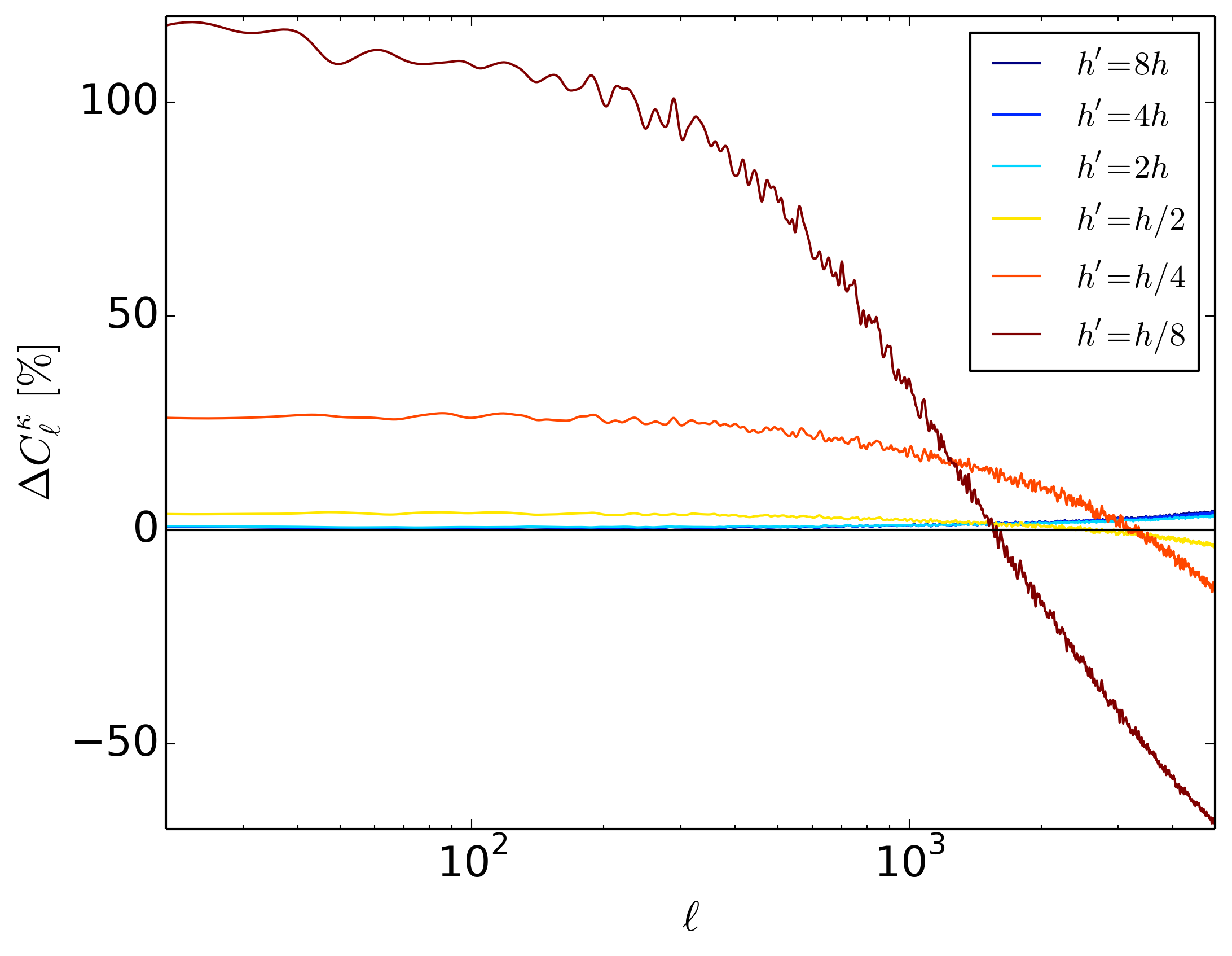} 
    \caption{Relative difference on the convergence angular power spectrum depending on the derivation step used in the infinitesimal method (see \cref{subsubsec:infinitesimal_beams}) to compute the lensing distortion matrix at $z = 1.9$, using Healpix with $N\e{side} = 4096$. Using a larger derivation step considerably biases the estimation of the power spectrum, while smaller steps mostly affect the very small scales. Our choice for $h$ seems ideal as it is stable for $\ell < 10^3$ and should not lead to any large bias at smaller scales.}
    \label{fig:reldiff_derivation_step_infinitesimal}
\end{figure}

The Laplacian of the gravitational potential along the line of sight in \cref{eq:infinitesimal_matrix} is computed using finite differences of the force. To compute these differences, we need a derivation step $h$ in  \cref{eq:laplacian_finite_difference}, which is set to the size of the most refined cell the photon is in. In \cref{fig:reldiff_derivation_step_infinitesimal} we show the impact of the derivation step on the estimation of the convergence angular power spectrum. 
We multiplied the size of our step choice by a factor 8, 4, 2, 1/2, 1/4, and 1/8 to clearly test our method. For a derivation step smaller than our reference choice, we see that there is a large bias (several orders of magnitude larger than in \cref{subsec:nsteps_per_amr_cell}) on the angular power spectrum even at linear scales (small $\ell$), and then at damping at small scales (high $\ell$). For derivation steps larger than the reference one, there is no noticeable difference at $\ell < 10^3$, while at $\ell > 10^3$ the angular power spectrum seems to be over-estimated with increasing $h'$. This overall behaviour shows that we find a `sweet spot' when setting $h = 2^{\rm level}$. This choice was also shown to lead to very good agreement with theoretical predictions from CLASS \citep{lesgourgues2011cosmic} in \cite{rasera2021raygal}.

\subsection{Mean convergence and the Born approximation}

Here, we estimate the impact of the commonly used Born approximation on weak-lensing quantities, and more precisely on the mean convergence. A standard test of the Born approximation with respect to real ray-tracing is to compute the convergence angular power spectrum and estimate the differences in both cases. In this particular configuration, the Born approximation is known to give results that are extremely close to real ray tracing \citep{hilbert2020accuracy}, but this should be taken with caution as it can be deceiving when considering galaxy surveys. Indeed, to compute the angular power spectra, it is common to propagate light rays in the direction of homogeneously distributed pixels on a map (or plane). However, this procedure gives the same weight to each pixel, which is not what really happens in galaxy surveys where solid angles on the sky are magnified. This is the difference between direction-averaging and source-averaging \citep{kibble2005average,bonvin2015cosmological,kaiser2016bias,breton2021theoretical}. When propagating light rays in random (or statistically isotropic) directions, it is expected that the Born approximation should give overall similar results to real ray tracing \citep{breton2021theoretical}. In particular, in both cases, we expect $\ev{\kappa} = 0$ (which also implies $\ev{\mu} > 1$ at second order), because photons evenly sample the sky. However, when we compute the null geodesics between the observer and sources, it is known that $\ev{\mu} = 1$ \citep{weinberg1976apparent} and $\kappa < 0$. Indeed, photons tend to propagate in more  under-dense regions between two fixed points for several realisations of the matter density field. 

\begin{figure}
\includegraphics[width=\columnwidth]{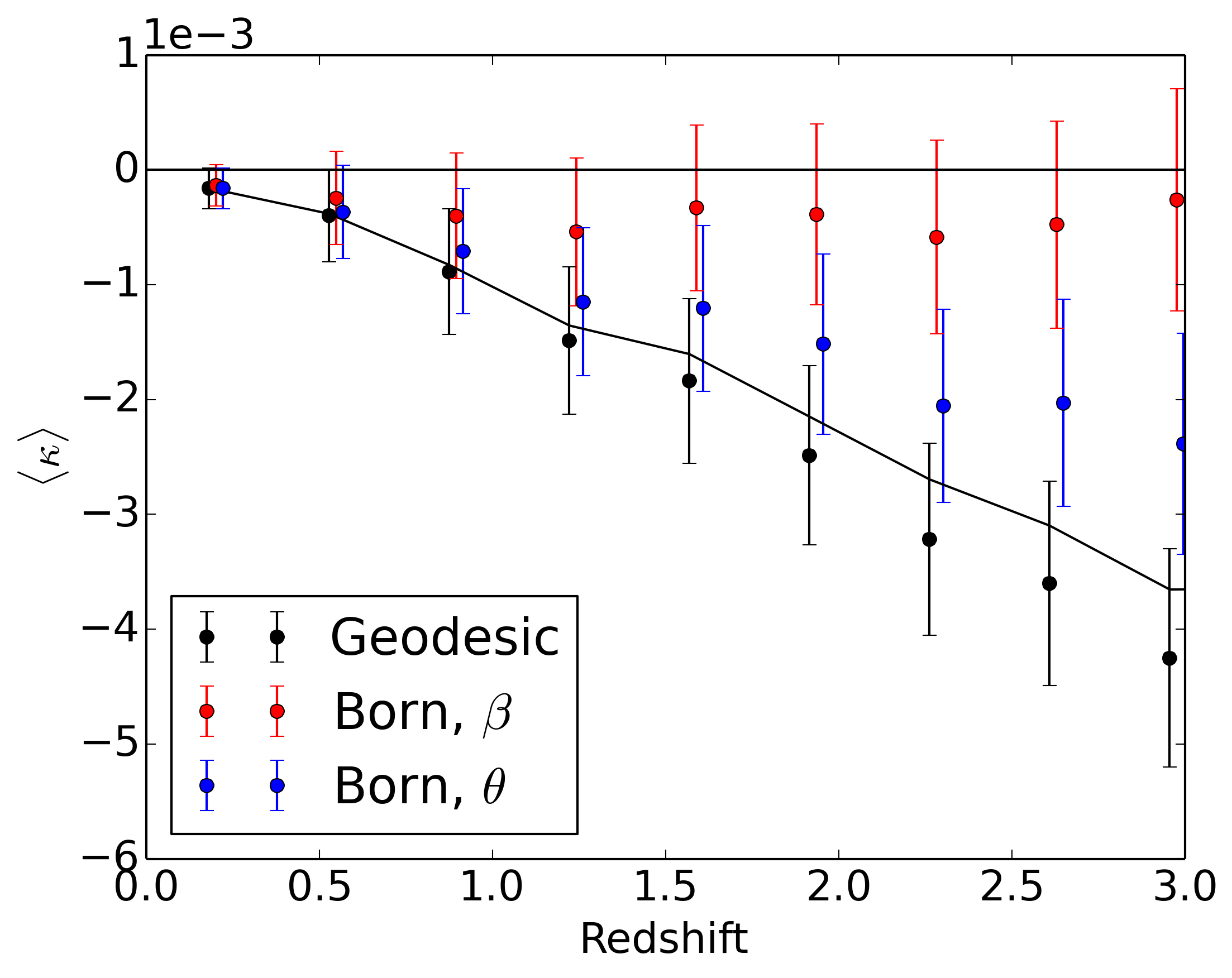} 
    \caption{Mean convergence on the deep narrow cone of RayGal (400 deg$^2$ aperture) as a function of redshift, when evaluating the distortion matrix along null geodesics (black) or using the Born approximation in the direction of sources ($\bm{\beta}$, red) or images ($\bm{\theta}$, blue). The black solid line shows the expected value of the mean convergence when using null geodesics, that is $\ev{\kappa} = \frac{1}{2}(\ev{\mu}-1) - 2\ev{\kappa}^2$, where the averaged quantities are numerically evaluated. The error bars contain Poisson and super-sample variance \citep{breton2021theoretical}. The red and black points are consistent with the expectation values for direction and source averaging procedures, respectively. The blue points, which come from a hybrid prescription, seem to lie between the former two.}
    \label{fig:meankappa}
\end{figure}

In \cref{fig:meankappa}, we compute the distortion matrix for a given simulated halo catalogue and estimate the averaged convergence in tomographic redshift bins.
In any case, we use the infinitesimal method in \cref{subsubsec:infinitesimal_beams}, and distinguish three different types of trajectories: first we consider the null geodesics between the observer and sources; then we use the Born approximation where photons propagate in straight lines between the observer and the sources or the source images. We note that the last case is strictly theoretical, because to the best of our knowledge it is not necessarily used nor discussed, but is interesting from a pedagogical point of view to cover all the possible scenarios. We see that for the standard Born approximation, that is when photons propagate in straight lines in the source direction ($\bm{\beta}$), the average convergence is consistent with zero within error bars, which is expected. For null geodesics, we find that the mean convergence is indeed negative and closely follows the theoretical prediction. Interestingly, when we use the Born approximation in the direction of images, the result seems to lie in between the first two cases.

There are two main implications of this result: first, a precise theoretical modelling of weak-lensing observables might need to account for the fact that $\ev{\kappa} \neq 0$ for galaxy surveys. Secondly (and more importantly), one should take the distortion matrix evaluated using the Born approximation  with caution when constructing realistic simulated catalogues, because in this case, the mean magnification is superior to unity (while in reality it should be strictly equal before the flux cut), and furthermore the magnification probability density function is that of the null-geodesic case multiplied by $\mu$ (this was also discussed in \citealt{takahashi2011probability}). When emulating a flux-limited survey, one has to magnify the fluxes (or magnitude) with the magnification; however, if one uses the Born approximation, the final number count on the flux-limited sample will be larger than it should realistically be in comparison to observations. 

\subsection{Relativistic effects and galaxy clustering analyses}

Last, we discuss the importance of the global relativistic treatment for light propagation. While ray-tracing codes usually only allow gravitational lensing studies, \codename{} offers a unified framework which accounts for both gravitational lensing and redshift perturbations. For the latter, we implemented all the corrections at first order in metric perturbations (as seen in \cref{subsec:catalogues}). This allows in particular to study the impact of relativistic effects for galaxy clustering analysis. The full relativistic number count was analytically estimated within linear theory by \cite{yoo2009new}, \cite{bonvin2011what}, and \cite{challinor2011linear}, and it was shown by \cite{bonvin2014asymmetric} that the dipole of the correlation function could be very sensitive to the gravitational potential and could therefore be a useful probe to test the nature of gravity \citep{bonvin2018testing}. \cite{breton2019imprints} found a nice agreement at large scale with linear-theory-based predictions, which shows the accuracy of our method. However, it was noted that the small scales are completely dominated by the gravitational potential, which is far beyond the expectation from linear theory. This was later analytically modelled using non-linear prescriptions \citep{didio2019relativistic,saga2020modelling,beutler2020modeling,saga2022detectability}. The dipole of the correlation function of galaxies is expected to be detected with a high signal-to-noise ratio for next-generation galaxy surveys \citep{saga2022detectability}, for which it will be mandatory to perform a full relativistic treatment of light propagation. 

\section{Conclusion}
\label{sec:conclusion}

In this paper, we present \codename{}, a ray-tracing framework which post-processes numerical simulations to accurately rebuild the past light cone of an observer from linear to non-linear scales. This framework propagates light rays in the AMR structure of a simulation light cone by solving the null geodesic equations of a perturbed FLRW metric in the Newtonian gauge and does not resort to any further approximation. Moreover, \codename{} is optimised for HPC and has already been run on very large $N$-body simulations such as DEUS- Full Universe Run \citep{alimi2012deus} and RayGal \citep{rasera2021raygal}. 

Our code produces relativistic simulated catalogues (where the null geodesic between the observer and sources are identified), Healpix maps, and various light-ray statistics along their propagation. By accounting for all the effects at first order in metric perturbations, \codename{} opens up a wide range of possible applications, some of which have already been studied in the literature:
\begin{itemize}
    \item \emph{Weak gravitational lensing}: \cite{rasera2021raygal} studied the impact of relativistic effects on various lensing-matter angular power spectra. In particular, they emphasised the importance of peculiar velocities and magnification bias beyond linear order. Furthermore, the weak gravitational lensing properties of light beams with finite extension is different from those of an infinitesimal beam. This effect was successfully confronted to theoretical predictions for the convergence and shear angular power spectra in \cite{breton2021theoretical}.
    \item \emph{Galaxy clustering and relativistic effects}: The simulated catalogues produced by \codename{} contain both the comoving and apparent angular position, as well as the full redshift decomposition for a given source. From this, one can perform galaxy clustering analysis beyond standard assumptions (distant observer, redshift only perturbed by peculiar velocities, etc.). In particular, in \cite{breton2021impact} the authors studied the impact of gravitational lensing (also known as magnification bias) on the estimation of the growth rate of structure when performing a standard RSD analysis and found that if lensing is not accounted for in the modelling, this leads to an underestimation of $f\sigma_8$. These catalogues also enabled for the first time the study of relativistic effects on the dipole of the correlation function at all scales \citep{breton2019imprints, taruya2020wide, saga2020modelling}, which will be detectable in next-generation surveys with high signal-to-noise ratios \citep{saga2022detectability}. Using the same data, \cite{beutler2020modeling} performed a similar analysis on the power spectrum dipole. 
    \item \emph{Distance measures}: Distance measures are crucial to interpreting observational data and cosmological inference. When it comes to the Hubble diagram, the average distance is often modelled by that of an FLRW universe. \cite{breton2021theoretical} carried out a detailed study of the perturbations on the distance--redshift relation for a wide range of redshifts. They numerically tested Weinberg's conjecture \citep{weinberg1976apparent}, which states that the area of constant redshift is unaffected by inhomogeneities in the matter density field and showed that even a full non-linear treatment gives similar results to the theoretical predictions of \cite{kaiser2016bias} regarding possible biases.
    \item \emph{Integrated Sachs-Wolfe effect}: \cite{adamek2019raytracing} used \codename{} on the DEUS- Full Universe Run simulations up to $z = 25$ to study the ISW angular power spectra for various scenarios such as $\Lambda$CDM, phantom dark energy, and modified gravity. In any case, the authors found good agreement with theoretical predictions.
\end{itemize}

These works show the importance of an exhaustive modelling of the light cone in order to correctly interpret future surveys, which will probe the Universe at the largest scales with unprecedented precision. In particular, with increasing data accuracy, omissions in the model might lead to biases in the inference of cosmological parameters. 
The methods developed in \codename{} will be useful, in coordination with theoretical predictions and observations, for refining existing probes and constructing new ones in order to shed light on the true nature of our Universe.

\begin{acknowledgements}
We thank Yann Rasera for helpful comments on the draft, and the Laboratoire Univers et Théories for its support and for providing a stimulating scientific environment from the early development of \libname{} to the last raytracing analyses with \codename{}. The design, development, and implementation of the foundational \libname{} library as well as the initial version of \codename{} were carried out at LUTH during VR's PhD under the direction of Jean-Michel Alimi and Yann Rasera. The further developments of \codename{} were carried out during MAB's thesis under the direction of Yann Rasera. We thank Jean-Michel Alimi and Yann Rasera for their scientific contributions and LUTH for its hospitality. MAB also thanks Julian Adamek for an early implementation of Healpix in the \codename{} code. VR also thanks the countless dedicated developers of the software stack without which none of this would have been possible. This work was granted access to HPC resources of TGCC/CINES through allocations made by GENCI (Grand Équipement National de Calcul Intensif) under the allocation 2020-A0070402287.
\end{acknowledgements}

\bibliographystyle{aa}
\bibliography{biblio} 

\end{document}